\documentclass[10pt,twocolumn,preprintnumbers,amsmath,amssymb,nofootinbib
,superscriptaddress]{revtex4}
\usepackage{graphicx,longtable,mathrsfs,color,array}
\usepackage[hidelinks]{hyperref}
\usepackage[usenames,dvipsnames]{xcolor} 
\usepackage{amssymb,amsmath,mathtools,mathrsfs,slashed} 
\usepackage{epsfig,subfigure,placeins,float} 
\usepackage{booktabs,longtable,ctable,multirow} 
\usepackage{exscale,relsize} 
\usepackage[normalem]{ulem} 
\usepackage{enumerate}
\usepackage{times, mathptmx} 
\usepackage{color}
\usepackage{hyperref}
\usepackage{graphicx}
\usepackage{color}
\usepackage{graphicx,graphics}

\newcommand{\be}{\begin{equation}}  
\newcommand{\ee}{\end{equation}}  
\newcommand{\bear}{\begin{eqnarray}}  
\newcommand{\eear}{\end{eqnarray}}  
\newcommand{\ba}{\begin{array}}  
\newcommand{\ea}{\end{array}}

\newskip\humongous \humongous=0pt plus 1000pt minus 1000pt  
\def\caja{\mathsurround=0pt}  
\def\eqalign#1{\,\vcenter{\openup1\jot \caja  
	\ialign{\strut
\documentclass[10pt,twocolumn,preprintnumbers,amsmath,amssymb,nofootinbib,superscriptaddress]{revtex4} \hfil$\displaystyle{##}$&$  
	\displaystyle{{}##}$\hfil\crcr#1\crcr}}\,}  
\newif\ifdtup

\def\oldreffmt#1{\rlap{[#1]} \hbox to 2\parindent{}}

\def\figfmt#1{\rlap{Figure {#1}} \hbox to 1in{}}  
  
\def\ie{\hbox{\it i.e.}{}}	\def\etc{\hbox{\it etc.}{}}  
\def\eg{\hbox{\it e.g.}{}}	  
\def\etal{\hbox{\it et al.}}  
  
\def\beq{\begin{equation}}  
\def\eeq{\end{equation}}  
\def\bea{\begin{eqnarray}}  
\def\eea{\end{eqnarray}}  
\def\half{\frac{1}{2}}  
  
\def\bq{\begin{quote}}  
\def\eq{\end{quote}}

\def\half{\frac{1}{2}}       

\def \etal {{\it et al.}\ }  
\relax  

\newdimen\tdim  
\tdim=\unitlength

\begin{document}

\preprint{FERMILAB-PUB--16-458-T}

\title {Weyl Current,
Scale-Invariant Inflation\\
and Planck Scale Generation 
}

\author{Pedro G. Ferreira}
\email{pedro.ferreira@physics.ox.ac.uk}
\affiliation{Astrophysics, Department of Physics\\
University of Oxford, Keble Road\\
Oxford OX1 3RH\\$ $}
\author{Christopher T. Hill}
\email{hill@fnal.gov}
\affiliation{Fermi National Accelerator Laboratory\\
P.O. Box 500, Batavia, Illinois 60510, USA\\$ $}
\author{Graham G. Ross}
\email{g.ross1@physics.ox.ac.uk}
\affiliation{Rudolf Peierls Centre for Theoretical Physics, \\
University of Oxford, 1 Keble Road\\
Oxford OX1 3NP\\$ $}

\date{\today}

\begin{abstract}

Scalar fields, $\phi_i$, can be coupled non-minimally
to curvature and satisfy the general
criteria: (i) the theory has no mass input parameters, including $M_{P}=0$; 
(ii) the $\phi_i$ have arbitrary values and gradients, but undergo
a general expansion and relaxation to constant values that satisfy
a nontrivial constraint, ${K}(\phi_i)=$ constant; (iii) this
constraint breaks scale symmetry spontaneously, and the Planck mass is dynamically 
generated; (iv) there can be adequate inflation
associated with slow roll in a scale invariant potential subject to the constraint; 
(v) the final vacuum can have a small to vanishing
cosmological constant  (vi) large hierarchies in VEV's can naturally form; (vii) there is a 
harmless dilaton
which  naturally eludes the usual constraints on massless scalars.  
These models are governed by a global Weyl scale symmetry and its conserved current, $K_\mu$.
At the quantum level the Weyl scale symmetry
can be maintained by an invariant specification of renormalized quantities.
\end{abstract}

\maketitle

\section{Introduction}


There has recently been considerable interest in scale symmetric general
relativity, in conjunction with inflation and dynamically generated 
mass scales \cite{ShapoZen,ShapoZen2,ShapoBlas,ShapoBell,Linde,All,JP,Kann,Kar,ShapoKar,Quiros,FHR,Kannike}. 
This is a theory containing fundamental scalar fields together with
general covariance and non-minimal coupling of
the scalars to curvature, but no Planck mass.   Remarkably, starting with a scale invariant
action, it is possible to spontaneously generate the Planck mass scale itself 
and naturally produce significant inflation.
The inflation can, moreover, lead to large hierarchies of
scalar vacuum expectation values (VEV's).  
All of this occurs as one unified phenomenon. 

The key ingredient of this mechanism  is a {\em global} Weyl scale symmetry and
its current, $K_\mu$.  Gravity
drives  the scale current density, $K_0$, to zero, much as any conserved current charge density
dilutes to zero by general expansion.  However, the particular
structure of the  $K_\mu$ current is such that it has a ``kernel,'' \ie, $K_\mu =\partial_\mu K$.  Hence, as
the scale charge density is diluted away, 
$K_0\rightarrow 0$,  the kernel evolves as  $K\rightarrow $ constant.  $K $
is the order parameter that defines a spontaneous scale symmetry breaking and the Planck scale,
$K=O(M_{P}^2)$.
The breaking of scale symmetry here is ``inertial,'' and is determined by the  
random initial values of the field VEV's that settle
down to yield a random fixed value of $K$.   

In the multi-field case the role of the potential is to determine 
the relative VEV's of the scalar fields contributing to $K$.
In this case the nonzero constant value of $K$ defines a constraint on the scalar field VEV's, 
requiring that the VEV's lie on an  ellipse in  multi-scalar-field space.
 The inflationary slow-roll conditions are consistent with constant $K$ and an inflationary era readily occurs  
in which the field VEV's migrate along the ellipse,   and
ultimately flow to an infra red fixed point. 
For the special case that the potential has a flat direction the fixed point corresponds
to the potential minimum, the field VEV's flow to it, and the final cosmological constant vanishes.

In the present paper we discuss how this ``current algebra'' works in detail, and how inflation and
Planck scale generation emerge from it.
We will first illustrate this phenomenon in Section II,
in a simplified theory with a single scalar field, $\phi $, and a non-minimal coupling to gravity
$\sim -(1/12)\alpha\phi^2 R$. For us $\alpha<0$, and  a nonzero VEV of $\phi$
induces a positive Planck (mass)$^2$. We allow scale invariant potentials, such as
$\lambda \phi^4$.    This theory thus has a global Weyl scale
symmetry, and a conserved scale current:
\beq
 K_\mu =(1-\alpha)\phi \partial_\mu \phi.
\eeq
The prefactor is relevant and nontrivial when we consider $N$ scalar fields
(this current vanishes in the $\alpha=1$ limit when the Weyl symmetry becomes local \cite{JP}).

The Weyl scale current  kernel is,  $K=(1-\alpha)\phi^2/2$.
 The kernel, $K$, is driven to a
constant during an initial period of expansion of the universe, as $K_{0}$ is
diluted to zero. There is no ellipse in the single
field case, and the field comes to rest with a fixed, eternal VEV,
$\phi = \sqrt{2K/(1-\alpha)}$ .
The theory acquires the Planck mass as $M_P^2=-\alpha K/6(1-\alpha)$. 
and the resulting inflation is eternal.

The Nambu-Goldstone theorem applies with the dynamical spontaneous scale
symmetry breaking by nonzero $K$, and there is a dilaton.  We will mention
some of the properties of the dilaton, with a more detailed discussion
in  \cite{FHR1}. If the underlying
Weyl scale symmetry is maintained throughout the full theory
(including quantum corrections), then the massless 
dilaton has at most derivative coupling to matter and 
the  Brans-Dicke constraints 
go away. 

We discuss in Section III a model with two scalars, $\phi$ and $\chi$.  The generalization of
the Weyl current is straightforward. After the initial
expansionary phase establishing constant $K$,  the fields readily generate 
a period of slow-roll inflation as their VEV's  migrate along an ellipse 
defined by constant $K$.  
If the potential  $V(\phi_i)$ is scale invariant and has a nontrivial minimum with non-vanishing VEV's,  it follows
 that $V(\phi_i)$ vanishes at its minimum and
that it has a flat direction corresponding to a definite ratio of the scalar field VEV's. 
The slow-roll inflationary period is terminated by a 
period of ``reheating'' in which the fields acquire large kinetic energy which 
is rapidly damped by expansion. 
Subsequently the fields flow  toward
an infrared (IR) fixed point that determines the ratio of their VEV's in terms 
of the couplings appearing in the scalar potential (this was studied in a two 
field example in ref.\cite{FHR}). The fixed
point is the intersection of the potential flat direction with the ellipsoid. 
If the potential does not have a non-trivial minimum, gravitational effects prevent the roll 
to the scale invariant minimum and the inflation
is eternal, \ie, there is then a relic cosmological constant.

In Section IV we discuss the $N$-scalar scheme
and the analytic solution for the inflationary
phase in the two scalar scheme. We consider  generalized inflationary fixed point 
of the $N$ scalar schemes, 
and the $N=3$ model is examined
in detail.  

If scale symmetry is broken through quantum loops,
the resulting trace anomaly implies that $K_\mu$ is no longer
conserved. Then the field VEV's, hence $K$, would relax to zero, and with it would go the Planck mass.
To avoid this it is necessary to maintain the Weyl symmetry throughout. 
One of our main theses is that this
is possible, \ie, the Weyl symmetry can be maintained at the quantum level
{\em if no external mass scales are introduced into the theory during the 
process of renormalization}. 

In Section V we turn to the quantum effects. We first describe how the Einstein 
and Klein-Gordon equations are conventionally modified by scale anomalies,
leading to the modifield $K_\mu$ current and
the kernel $K$.  Our main goal here is to describe and construct effective
Coleman-Weinberg-Jackiw \cite{CW,Jackiw10} actions where the couplings
run with fields.      

In Weyl invariant theories there can be no absolute meaning to mass;
only Weyl invariant dimensionless ratios of mass scales will occur.
It is  therefore crucial that no ``external
mass scales'' are introduced at the quantum level in 
renormalizing the theory.  This implies that  counterterms must
be field dependent and are ultimately specified by the overall constraint
that the renormalized action remains Weyl invariant. In the effective
action the running couplings must therefore depend exclusively upon 
Weyl invariant ratios of values of field VEV's,
\eg, $\lambda(\phi_c/\chi_c)$, rather than ratios involving
 some external mass scale, \eg, $\lambda(\phi_c/M)$. 
This approach  makes no specific reference to any particular
regularization method (see \cite{Percacci,bardeen}). The renormalization group 
with nontrivial $\beta$-functions remains, however the running
of parameters, is now given in terms of Weyl invariants.   

{We give general formal arguments in Section V
and more details will be given elsewhere \cite{FHR2}. 
In Section V we explore a simple two scalar model of quantum effects with a particular
choice of the running renormalized couplings which are expected to emerge in
detailed calculations. Since the renormalization group  running occurs in Weyl invariants
such as $\phi_c/\chi_c$ rather than $\phi_c/M$,  we find that the ellipse can be significantly
distorted by these effects. $K$  becomes constant, and a
nontrivial ratio  of VEV's $\phi_c/\chi_c$ develops which is  suggestive that
 a hierarchical relationship between $M_P$, $M_{GUT}$ and $m_{Higgs}$ 
might emerge from this dynamics in more detailed models.
We follow with conclusions.    }

\section{Single Non-Minimal Scalar}

\subsection{The Action}

We begin by establishing some notation.
A standard Einstein gravitation in our sign conventions with a minimally coupled massless scalar field,
${\sigma}$, and metric tensor ${g}$
and cosmological constant, $\Lambda$,
is an action of the form:\footnote{Our metric signature convention is $(1,-1,-1,-1)$, and our sign
convention for the Riemann tensor is that of Weinberg \cite{Weinberg}; our conventions are 
identically those of reference \cite{CCJ}.}
\bea
\label{zero}
S=\int \sqrt{-{g}}\left( \frac{1}{2}{g}^{\mu \upsilon }\partial _{\mu }{\sigma}
\partial _{\nu }{\sigma} -\Lambda +\frac{1}{2}M_P^{2}R\right) 
\eea
where the Einstein-Hilbert term contains the scalar curvature, $R$, and
the Planck mass: $M_P^{2}=\left( 8\pi G\right)^{-1}$.  For small ${\sigma}$
this action describes a deSitter universe with Hubble parameter:
\beq
\label{zeroH}
 H^2 = \frac{\Lambda}{3M_P^2}
\eeq

Presently, we consider a theory of a real scalar field, $\phi$, in which
the Einstein-Hilbert term has been replaced with the non-minimal scalar coupling
$-({1}/{12})\alpha \phi^{2}R$,  and we choose a scale invariant potential $V(\phi) = {\lambda}\phi^4/4 $:
\bea
\label{one}
S=\int \sqrt{-g}\left( \frac{1}{2}g^{\mu \upsilon }\partial _{\mu }\phi
\partial _{\nu }\phi -\frac{\lambda}{4}\phi^4 -\frac{1}{12}\alpha \phi
^{2}R\right) 
\eea
Assuming $\phi$  acquires a VEV,
we would generate a Planck mass from eq.(\ref{one}) of the form
$M_P^2 = -\alpha \phi^2/6$. We thus require $\alpha <0$, to obtain
the correct-sign for the  Einstein-Hilbert term, as in eq.(\ref{zero}).

The theory of eq.(\ref{one}) is globally scale invariant.
The invariant scale transformation  corresponds to the global limit of the
``Weyl transformation:''
\bea
\label{Weyl}
g_{\mu\nu} \rightarrow e^{-2\epsilon(x)}g_{\mu\nu} \qquad \phi \rightarrow e^{\epsilon(x)}\phi(x)
\eea
with $\epsilon(x)=\epsilon$ being constant in space-time. 
If we perform an infinitesimal local transformation as in eq.(\ref{Weyl}) on the action, we obtain the
Noether current:
\bea
\label{current}
K_\mu = \frac{\delta S}{\delta \partial^\mu \epsilon}=(1-\alpha)\phi\partial_\mu\phi\;.
\eea
For the $N=1$ single scalar case the prefactor of $(1-\alpha)$ appears spurious,
but it is an essential normalization for $N> 1$ scalars where the $\alpha_i$ can
take on different values, and this factor
is generated when the Noether variation is performed, and it describes the
vanishing of $K_\mu$ in the limit $\alpha \rightarrow 1$ which
corresponds to a particular locally Weyl invariant theory \cite{JP}.

The existence and conservation of $K_\mu$  follows by use of the equations of motion. 
From eq.(\ref{one}) we obtain the Einstein equation:
\bea 
\label{einstein1}
\frac{1}{6}\alpha \phi ^{2}G_{\alpha \beta } & = &
\left(\frac{3-\alpha}{3} \right)\partial _{\alpha }\phi \partial _{\beta }\phi
-g_{\alpha \beta }\left( \frac{3-2\alpha}{6}\right) \partial
^{\mu }\phi \partial _{\mu }\phi 
\nonumber \\
& & \!\!\!\!\!  \!\!\!\!\! \!\!\!\!\! \!\!\!\!\!     
+\frac{1}{3}\alpha \left( g_{\alpha \beta
}\phi D^{2}\phi -\phi D_{\beta }D_{\alpha }\phi \right) +g_{\alpha \beta
}V(\phi) 
\eea
The trace of the Einstein equation becomes:
\bea 
\label{trace1}
\!\!\!\!\!\!\!\!\!\! \!\!\!\!\!\!
-\frac{1}{6}\alpha \phi ^{2}R & = &  (\alpha-1 )\partial ^{\mu }\phi
\partial _{\mu }\phi +\alpha \phi D^{2}\phi +4V(\phi)
\eea
We also have the Klein-Gordon (KG) equation for $\phi$:
\bea
\label{KG1}
0=\phi D^{2}\phi+\phi\frac{\delta }{\delta \phi }V\left( \phi
\right) +\frac{1}{6}\alpha \phi^2 R
\eea
We can combine the KG equation, eq.(\ref{KG1}), and trace equation, 
eq.(\ref{trace1}), to eliminate the $ \alpha \phi^2 R $ term, and obtain:
\bea
\label{Kdiv00}
0& =& (1-\alpha )\phi D^{2}\phi +(1-\alpha )\partial ^{\mu }\phi \partial
_{\mu }\phi 
\nonumber \\
& & \qquad\qquad +\phi \frac{\partial }{\partial \phi }V\left( \phi \right) -4V\left(
\phi \right) 
\eea
This can be written as a current divergence equation:
\bea
\label{Kdiv1}
D^\mu K_\mu = 4V\left(
\phi \right) -\phi \frac{\partial }{\partial \phi }V\left( \phi \right) 
\eea
where $K_\mu$ is given in eq.(\ref{current}).
For the scale invariant potential, $V(\phi)\propto \phi^4$, the {\em rhs}
of eq.(\ref{Kdiv1}) vanishes
and the $K_\mu$ current is then covariantly conserved:
\beq
\label{cons1}
D^\mu K_\mu = 0.
\eeq
We emphasize that this is an ``on-shell'' conservation law, \ie, 
it assumes that the gravity satisfies eq.(\ref{einstein1}).

\subsection{The Kernel}\label{kkk3}

It is clear that the scale current can be written as $K_\mu=\partial_\mu K$ 
where the  kernel $K=(1-\alpha)\phi^2/2$.
This has immediate implications for the dynamics of this theory.
Consider  a Friedman-Robertson-Walker (FRW) metric:
\bea
g_{\mu\nu }& =&  [1,-a^{2}(t),-a^{2}(t),-a^{2}(t)] \qquad H=\frac{\dot{a}}{a}
\nonumber \\
G_{00}&= & -3\frac{\dot{a}^2}{a^2} \qquad R = 6\left( \frac{\overset{..}{a}}{a}+ \frac{\dot{a}^2}{a^2}\right) 
\eea  
Starting with an arbitrary classical $\phi$, after a period
of general expansion, in some regions of space 
 $\phi$ becomes approximately spatially constant, but time dependent.
{The conservation law of eq.(\ref{cons1}) becomes:}
\bea
\label{cons2}
\ddot K+3H \dot K=0
\eea
 If we take 
$\phi$ to be a function of time $t$ only, we have by eq.(\ref{cons2})
\begin{eqnarray}
K(t)=c_1+c_2\int^t_{t_{0}}\frac{dt'}{a^3(t')}.
\end{eqnarray}
where $c_1$ and $c_2$ are constants.
Therefore we find that, under general initial conditions,
$K(t)$ will evolve to a constant value, $K=K(t\rightarrow \infty)$.
The $(00)$ Einstein equation, with $G_{00}=-3H^2$, gives:
\beq
\label{fixed1}
 H^2=-\frac{\lambda\phi_0 ^{2} }{2\alpha }
\eeq
Thus, with $\alpha <0$ 
we have a self-consistent, exponential relaxation to constant  
 $\phi = \phi _0 
=\sqrt{2K/(1-\alpha)}$, 
and eternal inflation.

Note that this situation contrasts what happens in conventional Einstein gravity with a fixed
$M_{P}$ 
and a $\lambda \phi^4 /4 $ potential. Inflation is possible for super Plankian 
values of $\phi$ which slow-roll to $\phi = 0$.
Hence, while normal Einstein gravity causes $\phi$
to relax to zero, the  scale-invariant gravity
theory leads to constant nonzero $\phi=\phi_0$ which generates $M_{P}$ 
and eternal inflation.

Anticipating our discussion in Section V, we can ask how the trace anomaly,
arising through quantum effects,
would affect these conclusions? The Weyl current is
not conserved if there are trace anomalies, and eq.(\ref{Kdiv1}) becomes:
\bea
\label{Kdiv2}
D^\mu K_\mu = 4V\left(
\phi \right) -\phi \frac{\partial }{\partial \phi }V\left( \phi \right) =-\frac{\beta_\lambda(\phi)}{4}\phi^4
\eea  
where $\beta_\lambda (\phi)=d\lambda/d\ln\phi $ is the $\beta$-function associated with the 
radiative corrections of the quartic 
coupling $\lambda $ in eq.(\ref{one})\footnote{There is also an anomaly associated 
with the running of $\alpha$.}.  Indeed, this
anomaly enters the rhs of eq.(\ref{cons1}), and it would lead to slow-roll
relaxation of $\phi$ to zero, $K\rightarrow 0$, and thus the Planck mass
goes to zero as well. With non-zero trace anomaly, the enterprise of 
generating inflation and the Planck mass
as a unified phenomenon would then fail. One of our main arguments here is that 
we can maintain the Weyl symmetry in  any regularization scheme by renormalizing
the theory with counterterms that maintain
Weyl-invariance. $\beta$-functions then describe
the running of couplings in terms of Weyl invariants, such as $\beta_\lambda (\phi)=d\lambda/\ln(\phi/\sqrt{R})$,
but the trace anomaly is then zero, as discussed
in section V. This maintains the vanishing of the {\em rhs} of
eq.(\ref{cons1}), and the Planck mass is then stabilized.

\subsection{Weyl Transformation and the Dilaton}

We can identify the spatially constant field $\phi $ with a new field, $\sigma /f$
where $f$ is a ``decay constant'' (analogue of $f_\pi$), and $\phi_0$ is constant:
\beq
\phi =\phi _{0}\exp (\sigma/f ),
\eeq
and perform the metric transformation: 
\beq
g_{\mu \nu }=\exp (-2\sigma/f )\widetilde{g}_{\mu \nu }
\eeq
Using $g_{\mu \nu }=\exp (-2\epsilon)\widetilde{g}_{\mu \nu }$:
\beq
 R\rightarrow \exp (2\epsilon )\widetilde{R}+6\exp (2\epsilon 
)\left(  \partial ^{\mu }\epsilon   \partial _{\mu }\epsilon 
\ - \widetilde{D}^{\mu }\partial _{\mu }\epsilon  
\right) 
\eeq
where $\widetilde{R},(\widetilde{D}^{\mu })$ is the curvature 
(covariant derivative) expressed in terms of $\widetilde{g}_{\mu \nu }$,
we then have:
\bea
\label{dil}
S &=& \int \sqrt{-\widetilde{g}}\left[ \frac{%
\phi _{0}^{2}}{2f^2}\;\widetilde{g}^{\mu \nu }\partial _{\mu }\sigma 
\partial _{\nu }\sigma  -\frac{\lambda }{4}\phi _{0}^{4} \right.
\nonumber \\
& & \left.\!\!\!\!\! \!\!\!\!\! \!\!\!\!\!
-\frac{1}{2}%
\alpha \phi _{0}^{2}\left( \frac{1}{6}\widetilde{R}+\frac{1}{f^2}\partial^{\mu }\sigma
\partial_{\mu }\sigma   -\frac{1}{f}\widetilde{D}
^{\mu }\partial _{\mu }\sigma \right) \right] 
\eea
The canonical
normalization of the $\sigma$ field  thus requires
the decay constant $f= \sqrt{2K_0}$ where $K_0 =(1-\alpha)\phi_0^2/2$.
Dropping a total divergence,
and defining:
\beq
\Lambda =\frac{\lambda }{4}\phi _{0}^{4};\qquad \qquad M_P^{2}=-\frac{%
1}{6}\alpha \phi _{0}^{2}
\eeq
we have:
\bea
\label{fixed2}
S=\int \sqrt{-\widetilde{g}}\left( \frac{1}{2}\widetilde{g}^{\mu \nu
}\partial _{\mu }{\sigma } \partial _{\nu }{\sigma }
-\Lambda +\frac{1}{2} M_P^{2}\widetilde{R}\
\right) 
\eea
Therefore, we see that the scale invariant theory, 
 eq.(\ref{one}), can be viewed as the ``Jordan
frame,''
equivalent to the ``Einstein frame'' action eq.(\ref{fixed2}),
as we originally wrote down
in eq.(\ref{zero}). The massless  field 
${\sigma }$ is the dilaton, but this
feature is virtually hidden in the Einstein frame, since
there  $\sigma$ couples
to gravity only through it's  stress tensor.
Note the identical correspondence of eq.(\ref{fixed1}) 
with eq.(\ref{zeroH}).

Remarkably eq.(\ref{zeroH}) contains
a hidden Weyl symmetry.
We see that $\Lambda$ and
$M_P^2$ are related to the $\phi_0^2$,
and can be written in terms of the dilaton  decay constant as:
\beq
\Lambda =\frac{\lambda }{4(1-\alpha)^2}f^{4};\qquad M_P^{2}=-\frac{%
1}{6(1-\alpha)}\alpha f^2
\eeq
These relations are the analogue, 
in a chiral Lagrangian, of the Goldberger-Treiman relation,
$m_N=g_{NN\pi}f_\pi$ 
relating the mass of the nucleon, $m_N$, to $f_\pi$ and the strong
coupling constant $g_{NN\pi}$.
The variation of the action of eq.(\ref{fixed2})
with respect to $\sigma/f$
yields the current,  $K_\mu = f\partial_\mu\sigma $  which
 is the representation
 $K_\mu$  in the Einstein frame, and
the  analogue of the axial current, $f_\pi \partial_\mu \pi$, 
of the pion.

The dilaton reflects the fact that the
exact scale symmetry remains, though hidden in the Einstein frame.
We can rescale both the VEV $\phi_0 \rightarrow e^{\epsilon} \phi_0$
and the Hubble constant $H_0\rightarrow e^{\epsilon} H_0$ while their ratio remains fixed:
\beq
 \frac{ H_0^{2}}{\phi _{0}^{2}} = \frac{\lambda }{2\left| \alpha \right| }
\eeq 
It is straightforward to extend this effective Lagrangian to matter fields.
If the dilaton develops
a ``hard coupling'' to, \eg, the nucleon, then stars would
develop dilatonic halo fields. This would  then be subject to strict limits
from Brans-Dicke theories, and the models would fail to give acceptable inflation.
However, if all ordinary
matter fields have masses that are ultimately associated with the spontaneous breaking of the 
Weyl scale symmetry, then the dilaton only couples derivatively. There are then
no Brans-Dicke-like constraints, as no star or black-hole, etc., will generate
a $\sigma $ field halo. 
In a subsequent paper \cite{FHR2} we will discuss the dilaton phenomenology in greater detail.

\section{Two Scalar Theory}

\subsection{Classical Two Scalar Action}

Consider an
 $N=2$ model, with scalars $(\phi, \chi)$, and the potential:
\bea
\label{pot2}
W\left( \phi ,\chi \right) =\frac{\lambda}{4} \phi ^{4}+\frac{\xi}{4} \chi ^{4}+\frac{\delta}{2} \phi
^{2}\chi ^{2}
\eea

The action takes the form:
\bea
\label{action2}
S &= & \int \sqrt{-g}\left( \frac{1}{2}g^{\mu \upsilon }\partial _{\mu }\phi
\partial _{\nu }\phi +\frac{1}{2}g^{\mu \upsilon }\partial _{\mu }\chi
\partial _{\nu }\chi
\right.
\nonumber \\
& & \left.
 -W\left( \phi ,\chi \right) -\frac{1}{12}\alpha_1 \phi^{2}R-\frac{1}{12}\alpha_2 \chi ^{2}R\right) 
\eea
This has been studied in \cite{ShapoZen,ShapoZen2,ShapoBlas,ShapoBell,FHR}.
For example, \cite{ShapoZen2} study this theory in the context
of a unimodular gravity and perform a Weyl transformation taking  eq.(\ref{action2}) from a Jordan
frame to an Einstein frame. We follow the approach of \cite{FHR} and work directly
in the defining frame of eq.(\ref{action2}), and then just follow the dynamics.  The result is an
effective, emergent Einstein gravity where the Planck mass is induced by the VEV's of
$\phi$ and $\chi$.  We will see in Section IV that, due to the conserved $K$-current,
the slow-roll inflation of the classical system is amenable to an analytic treatment.
We will also extend this to include quantum corrections
that have a significant effect in the next section.

The sequence of steps follows those of the previous
single scalar case.  The Einstein equation is:

\scalebox{0.92}{\parbox{\linewidth}{
\bea
M_P^2 G_{\alpha \beta} 
&= &
\left(1-\frac{1}{3}\alpha_1 \right)\partial _{\alpha }\phi \partial _{\beta }\phi 
+\left(1-\frac{1}{3}\alpha_2 \right)\partial _{\alpha }\chi \partial _{\beta }\chi 
\nonumber \\
& &\!\!\!\!\! \!\!\!\!\! \!\!\!\!\! \!\!\!\!\! \!\!\!\!\! \!\!\!\!\! \!\!\!\!\!
-g_{\alpha\beta }\left( \frac{1}{2}-\frac{1}{3}\alpha_1 \right) \partial ^{\mu }\phi\partial _{\mu }\phi 
 -g_{\alpha \beta }\left( \frac{1}{2}-\frac{1}{3}\alpha_2 \right) \partial ^{\mu }\chi \partial _{\mu }\chi 
\nonumber \\
& &\!\!\!\!\! \!\!\!\!\! \!\!\!\!\! \!\!\!\!\! \!\!\!\!\! \!\!\!\!\! \!\!\!\!\!
+\frac{1}{3}\alpha_1 \left( g_{\alpha \beta }\phi D^{2}\phi -\phi D_{\beta }D_{\alpha}\phi \right) 
 +\frac{1}{3}\alpha_2 \left( g_{\alpha \beta}\chi D^{2}\chi -\chi D_{\beta }D_{\alpha }\chi \right) 
\nonumber  \\
& & 
\qquad \qquad +g_{\alpha \beta}W\left( \phi ,\chi \right) 
\eea
}}
where:
\bea
M_P^2=-\scalebox{0.95}{$\frac{1}{6}$}\left( \alpha_1 \phi ^{2}+\alpha_2 \chi ^{2}\right)
\eea
The trace of the Einstein equation becomes:
\bea
R & = & \frac{1}{M_P^2}\left(
(\alpha_1-1 )\partial ^{\mu }\phi\partial _{\mu }\phi+\left(\alpha_2-1 \right) \partial ^{\mu }\chi \partial _{\mu }\chi 
\right.
\nonumber \\
& & \qquad \left.
+\alpha_1 \phi D^{2}\phi +\alpha_2 \chi D^{2}\chi +4
W\left(
\phi ,\chi \right) \right)
\eea
The Klein-Gordon equations
for the scalars are:
\bea
0 & =& D^{2}\phi+
\delta \phi^{2}\chi+\lambda \phi^{3}+\frac{1}{6}\alpha_1\phi R
\nonumber \\
0 & =& D^{2}\chi+
\delta \phi\chi^{2}+\xi \chi^{3}+\frac{1}{6}\alpha_2\chi R
\eea
and we again use the trace equation to elimate $R$:
\bea
0 & = & \phi D^{2}\phi -\frac{\alpha_1 \phi ^{2}}{6M_P^2}\left( (1-\alpha_1 )\partial ^{\mu }\phi\partial _{\mu }\phi
+\left(1-\alpha_2 \right)\partial ^{\mu }\chi \partial _{\mu }\chi  \right.
\nonumber \\
& & \!\!\!\!\! \!\!\!\!\!
-\alpha_1 \phi D^{2}\phi -\alpha_2 \chi D^{2}\chi -4W\makebox{\huge $ \left. \right) $}
+\delta \phi^{2}\chi ^{2}+\lambda \phi ^{4}
\nonumber \\
0 & = & \chi D^{2}\chi -\frac{\alpha_2 \chi ^{2}}{6M_P^2}\left( (1-\alpha_1 )\partial ^{\mu }\phi\partial _{\mu }\phi
+\left( 1-\alpha_2\right)\partial ^{\mu }\chi \partial _{\mu }\chi \right.
\nonumber \\
& &  \!\!\!\!\! \!\!\!\!\!
-\alpha_1 \phi D^{2}\phi -\alpha_2 \chi
D^{2}\chi -4W \makebox{\huge $ \left. \right) $}  +\delta \phi^{2}\chi ^{2}+\xi \chi ^{4}
\eea
We again see that the sum of the Klein-Gordan equations
implies the conserved current, where the potential terms cancel owing to scale invariance:
\bea
0 & =&  D_{\mu }[(1-\alpha_1 )\phi \partial^{\mu }\phi +\left( 1-\alpha_2\right) \chi \partial^{\mu }\chi ]
\eea
so:
\beq
\label{Kcurrent2}
K_{\mu }=(1-\alpha_1 )\phi \partial^{\mu }\phi +\left( 1-\alpha_2 \right) \chi
\partial^{\mu }\chi 
\eeq
is conserved $D_{\mu }K_{\mu }=0$.  The kernel is now given by:
\beq
\label{kk2}
K=\half\left[(1-\alpha_1 )\phi^2 +\left( 1-\alpha_2 \right) \chi^2\right]
\eeq

\subsection{Synopsis of Two Scalar Dynamics}

\begin{figure}[tbp]
\vskip-0.1in
\begin{flushleft}
\includegraphics[width=8.9cm]{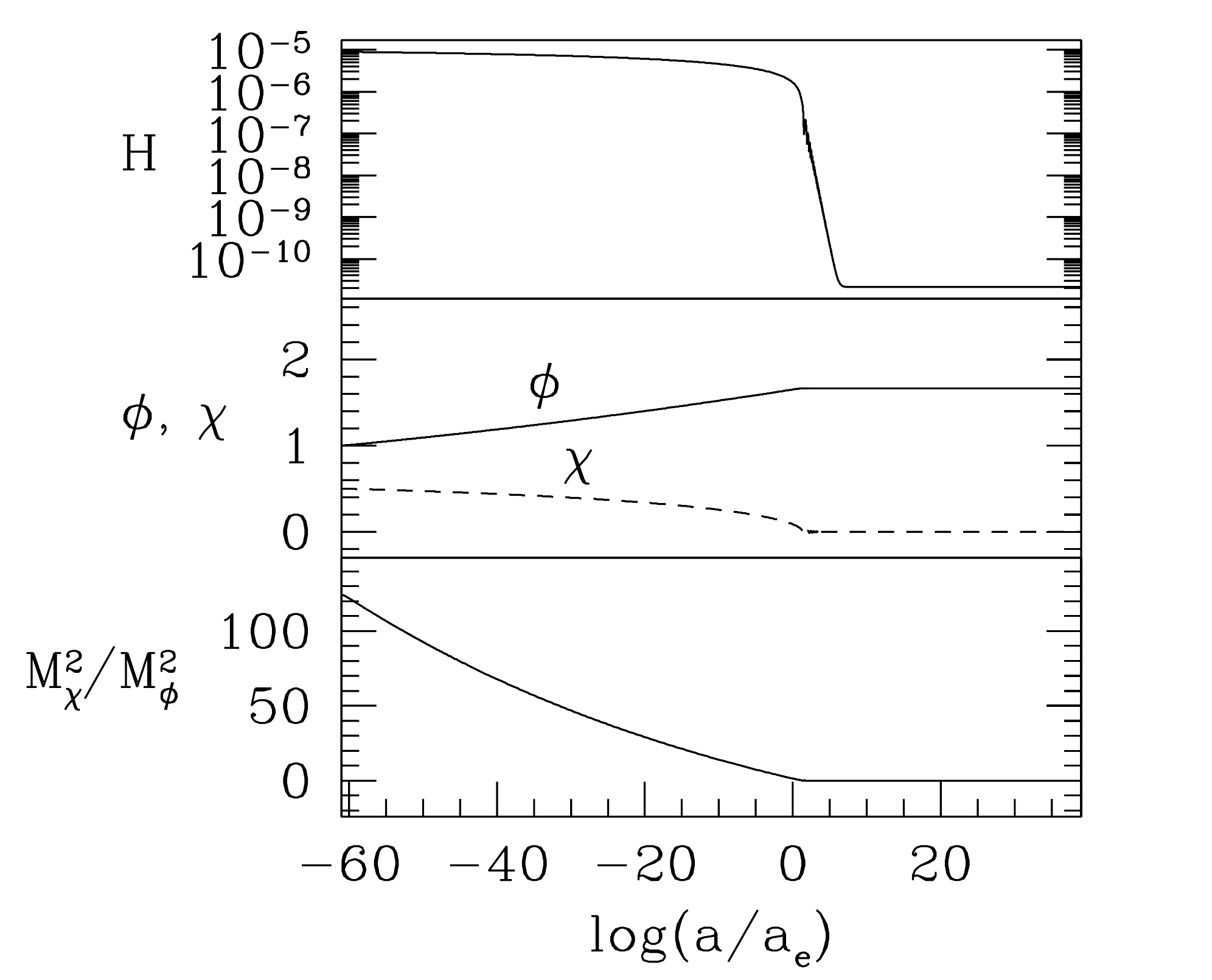} 
\end{flushleft}
\caption{Plot of the Hubble parameter, $H$, $\phi$, $\chi$ and 
the ratio of the two components of the effective Planck mass, $M^2_\phi$ and $M^2_\chi$,
as a function of $a$; we have normalized the x-axis to the scale factor at the end 
of inflation, $a_e$. The chosen parameters are:
$\alpha_1=-1.25\times 10^{-2}$; $\alpha_2=-6.19$;  $\delta=0$;
$\lambda=10^{-24}$; $\xi=10^{-9}$;
$\phi(-60)=1$ and $\chi(-60)=0.5$ in Planck units; 
initial velocities are set to zero.}
\label{figure_back}
\end{figure}

The two scalar theory has a number of interesting features, which we
will summarize presently. We first discuss the classical case
and, after the discussion of the scale invariant renormalization procedure, 
we consider the modifications that can occur when including radiative corrections in Section V. 

The potential of eq.(\ref{pot2}) has  the general form:
\bea
\label{pot3}
W\left( \phi ,\chi \right) =\frac{\xi}{4}\left(\chi^2 -\varsigma^2  \phi ^{2}\right)^2+{\lambda'\over 4}\phi^4
\label{pot1}
\eea
with $\lambda'=\lambda-\xi \varsigma^4$. For the case $\lambda'=0$ the potential has a flat direction 
with $\chi=\varsigma \phi$, and the vacuum energy vanishes 
for non-zero VEV's of the fields.

The theory can lead to a realistic cosmological evolution as illustrated in Fig.(\ref{figure_back}) 
 for a representative choice of parameters and initial conditions.
In an initial ``transient phase,'' the theory will redshift from arbitrary initial 
field values and velocities,  $(\phi, \dot{\phi}; \chi,\dot{\chi})$. 
Owing to the conserved $K$ current,
the redshifting will 
cause $(\dot{\phi},\dot{\chi})\rightarrow 0$ and the $K_0$ charge density to dilute away as $\sim a(t)^{-3}$ 
leading to   a state with constant kernel $K$.
The arbitrary, nonzero value of $K$, determines the scale of the Planck mass, $K\sim M_P^2$, and
 spontaneously breaks scale symmetry.  The fields $(\phi,  \chi)$
are now approximately constant in space VEV's and are constrained to lie
on the ellipse defined by eq.(\ref{kk2}). This initial location 
of the VEV's on the ellipse, $(\phi(0),  \chi(0))$,
is random.

As $K$ settles down to its constant value, Einstein
gravity has emerged with a fixed Planck mass.  
This can be seen analytically for
the classical case as in Section III.C below. 
The initial values of $(\phi_0,\chi_0)$ 
are random and would not
be expected to lie on the flat direction. 

\begin{figure}[t]
\includegraphics[width=8.9cm]{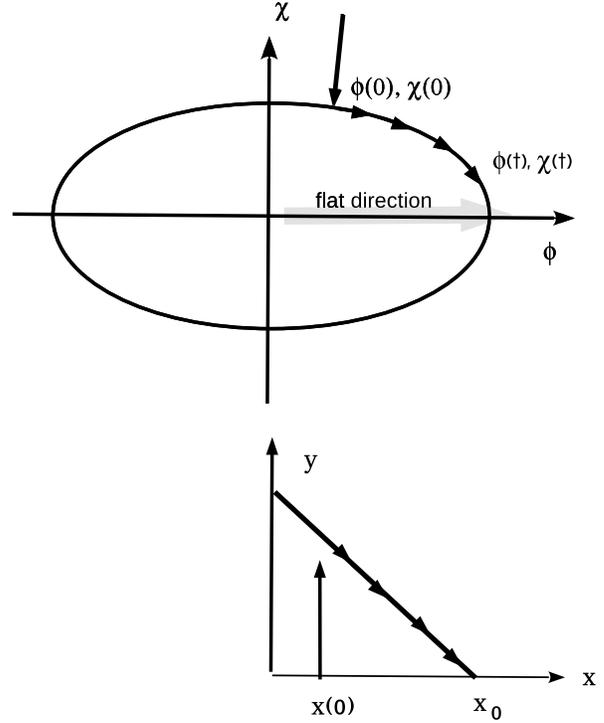} 
\vspace{-2cm}
\vspace{-0.0cm}
\caption[]{ The $K$-ellipse in $(\phi, \chi)$.  The potential flat direction
lies along the $\phi$ axis for the potential $\zeta \chi^4$.  Initial values of fields 
and their velocities rapidly redshift to constant $K$ and then  slow-roll on the ellipse
toward the fixed point. The ellipse is mapped into $x+y=1$;
initial value of $x=x(0)$ slow-rolls to the final fixed point $ x_0$ (presently $x_0 \approx 1$).}
\label{xroots}
\end{figure}

For a significant region of initial values the fields then slow-roll
along the ellipse, migrating toward a minimum of the potential and generating  a period of inflation. 
The flat direction is a  ray in the $(\phi,\chi)$ plane that intersects the 
ellipse defined by the kernel, eq.(\ref{kk2}).  If we assume $\varsigma<<1$ this intersection occurs
near the right-most  end of the ellipse where $\chi << \phi$
in quadrant I $(\phi,\chi)>0$ in Fig.\ref{xroots}.  {Note that $\varsigma<<1$ 
is a particular choice of the dynamics,
since for $\varsigma\sim 1$ the flat direction can be arbitrary in the $(\phi,\chi)$ plane, 
and the inflation
can still be significant, but we will not then
generate a large hierarchy in the VEV's of $\phi$ and $\chi$.}

The inflationary period ends when the slow-roll conditions are violated  and the system enters a period of ``reheating'' when the potential energy is converted to kinetic energy which rapidly redshifts. Although this period cannot be solved analytically a numerical simulation shows that the kernel remains constant and that the fields ultimately resume slow-roll with expectation values that are in the domain of attraction of an infra red fixed point
 \cite{FHR}\footnote{By coupling $\chi$  to standard model fields one has that energy will be transferred - the Universe will "reheat" - during the oscillatory phase; the oscillations will be damped driving the dynamics to the fixed point (which remains unchanged)}. The fixed point is 
determined by the parameters of the potential and
the $\alpha_i$ and, if the potential has a non-trivial minimum corresponding to $\lambda'=0$ in eq(\ref{pot1}),
the fixed point corresponds to the minimum of the potential with vanishing cosmological constant
(otherwise the fixed point corresponds to non-vanishing VEV's, with non-vanishing potential energy, leading to eternal inflation).

\subsection{Inflation in the Two Scalar Scheme}

In this section we give a detailed analysis of the inflationary era in the two scalar theory and determine the full  analytic solution in the slow-roll regime.  

In what follows we will be interested in  a large hierarchy between 
the scalar VEV's that can develop after an initial period of inflation. In this case the 
large field VEV (we will choose parameters such that this is the $\phi$ VEV) sets 
the magnitude of the Planck scale while the small field VEV sets the scale in the 
``matter'' sector characterised by the $\chi$ field.
\footnote{In \cite{ShapoBez,BezMagShapo} this field models the Higgs of the Standard Model.}


As before we assume the potential
of eq.(\ref{pot2}), and
that there is an Hubble size volume in which the fields are time dependent but spatially constant. 
Then following the argument in Section \ref{kkk3} we see that the kernel 
 $K$ becomes a constant, which we take to be an arbitrary
mass scale (related ultimately to the Planck mass
$K \sim  M_P^2$).
 The residual motion of the scalars during slow-roll is constrained to lie 
on the $K$ = constant ellipse and
is then described by a difference of the KG equations.
We thus form the convenient combination:
\bea
\label{KGcombo}
\alpha_2\frac{D^{2}\phi}{\phi} -\alpha_1 \frac{D^{2}\chi}{\chi} & = &
\nonumber \\
& &
\!\!\!\!\!  \!\!\!\!\! \!\!\!\!\! \!\!\!\!\! \!\!\!\!\!\!\!\!\!\! \!\!\!\!\! \!\!\!\!\!\!\!\!\!\! 
 -(\alpha_2\lambda-\alpha_1\delta) \phi ^{2} +(\alpha_1\xi-\alpha_2\delta)\chi ^{2}
\eea

We take the slow-roll limit of eq.(\ref{KGcombo})  
and we pass to the ``inflation derivative'' $D^2\phi
\rightarrow 3H\dot{\phi}=3H^2 \partial_N\phi$ where $N=\ln(a(t))$
hence $\partial_t\phi = H\partial_N\phi$:
\bea
\label{KGcombo2}
\frac{3}{2}H^2\left(\alpha_2 \frac{\partial_N\phi^2}{\phi^2} -\alpha_1  \frac{\partial_N\chi^2}{\chi^2} \right)
& = & 
\nonumber \\
& & \!\!\!\!\! \!\!\!\!\! \!\!\!\!\! \!\!\!\!\!\!\!\!\!\! \!\!\!\!\! \!\!\!\!\! \!\!\!\!\!
\!\!\!\!\! \!\!\!\!\! \!\!\!\!\! \!\!\!\!\! \!\!\!\!\! \!\!\!\!\! 
 -(\alpha_2 \lambda-\alpha_1 \delta) \phi ^{2} +(\alpha_1 \xi-\alpha_2 \delta)\chi ^{2}
\eea
We eliminate $H^2$ using the $(00)$ Einstein equation in the slow-roll
limit:
\bea
M_P^2G_{00} =-\half H^2(\alpha_1 \phi^2 +\alpha_2 \chi^2)\approx g_{00}W
\eea
Without loss of generality we can choose the ellipse $K = 1$ and we can map quadrant I 
of the ellipse into the variables: 
\beq
x=(1-\alpha_1  )\phi ^{2}\qquad \makebox{and}\qquad  y=(1-\alpha_2  )\chi ^{2}
\eeq
The ellipse then becomes the line segment $1=x+y$ in quadrant I.
With the fields constrained to be on the ellipse,
we see that $(x,y)$ are each constrained to range from $0$ to $1$ 

The slow-roll
differential equation on the ellipse,
eq.(\ref{KGcombo2}), can then be written as:
\beq
\label{slowroll}
\partial _{N}x=\frac{S(x)}{W(x)}x\left( 1-x\right) (x-x_{0}).
\eeq
where:
\bea
S(x) & = & \frac{2}{3}\frac{((1-\alpha_2  )A+(1-\alpha_1  )B)}{(1-\alpha_1  )^{2}(1-\alpha_2 
)^{2}}\times
\nonumber \\
& & \qquad
\frac{(\left( \alpha_1  -\alpha_2 \right) x+\alpha_2  (1-\alpha_1  ))}{\left( \alpha_2 
\left( 1-x\right) +\alpha_1  x\right) }
\eea
\bea
A & =& \left( \alpha_2  \lambda -\alpha_1  \delta \right)
\nonumber \\
B & =& \left( \alpha_1  \xi -\alpha_2  \delta \right) 
\eea
and: 
\beq
W\left( x\right) =\frac{\lambda x^{2}}{4(1-\alpha_1  )^{2}}+\frac{\xi
(1-x)^{2}}{4\left( 1-\alpha_2 \right) ^{2}}+\frac{\delta x(1-x)}{2(1-\alpha_1 
)\left( 1-\alpha_2 \right) }
\eeq
$x_0$ is the ``fixed point'' in $x$, as defined in \cite{FHR},
and takes the form:
\beq
x_{0}=\frac{B(1-\alpha_1  )}{%
A(1-\alpha_2  )+B(1-\alpha_1  )}
\eeq

The solutions to eq.(\ref{slowroll})  depend critically on the behaviour of ${S(x)}/{W(x)}.$
To demonstrate there is a region of parameter space that does undergo slow-roll inflation
we consider the case studied in ref.\cite{FHR}, in which $\xi\gg \delta\gg \lambda$, 
such that $B>>A,\;x_0\approx 1$ and, during the initial inflationary era,  $W \approx \xi(1-x)^2/
4(1-\alpha_2)^2$ . 
In this case:
\bea
\label{slowroll3}
 \partial _{N}x & =& -\frac{4}{3}x\frac{\alpha_1  }{(1-\alpha_1  )}\frac{(\left(
\alpha_1  -\alpha_2 \right) x+\alpha_2  (1-\alpha_1  ))}{\left( \alpha_2  +\left( \alpha_1 
-\alpha_2 \right) x\right) }
\nonumber \\
\eea

The above result is an exact solution for slow-roll in
the model of \cite{FHR}. 
The slow-roll conditions are readily satisfied for small, negative $\alpha_1 $ 
in which case:
\beq
\partial_N x\approx- {4\over 3}\alpha_1  x
\eeq
and $x(t)$  will roll from an initial $x(0)$ toward $x(t_E)=x_0\approx 1$
where $t_E$ is the time at the end of inflation.

Eq.(\ref{slowroll3})  can readily be integrated: 
\bea
\label{eq95}
 && \ln \frac{x(t)}{x(0)}-\alpha_1  \ln \left( \frac{\alpha_2  \alpha_1  -\alpha_1  x(t)-\alpha_2  (1-x(t))}{%
\alpha_2  \alpha_1  -\alpha_1  x(0)-\alpha_2 (1-x(0))}\right) 
\nonumber \\ 
&& \qquad\qquad\qquad =-\frac{4}{3}\alpha_1  (N(t)-N(0))
\label{srsoln}
\eea
In this limit of small $\alpha_1$
 eq.(\ref{srsoln}) implies the number of e-folds of inflation, $N$, is given by
\beq
N=N(t_{E})-N(0)={3\over 4\alpha_1 }\ln \left({x(0)\over x(t_E)}\right)
\eeq
Inflation ends when slow-roll ceases corresponding to  
the inflation parameter, $\epsilon$, approaching unity:
$\epsilon=-(1/2)({d\ln H^2/dN})\approx 1$.
This implies: 
\beq
{2\over 3}\left({2\alpha_1\over 1-x(t_E)}-{\alpha_1\over 1-x(t_E)+\alpha_1 /\alpha_2}\right )  x(t_E)\approx 1
\eeq
hence, when  $x(t_E)= 1-O(\alpha_1 )$.
The number of e-folds of inflation
is weakly governed by the initial value on the ellipse,  $x(0)$.
This is any value of order, but less than, unity, \eg, $x(0)\sim 0.5$, so to get large $(N(t_E)-N(0))$  
we require $\left| \alpha_1  \right| <<1.$

The resulting values for the spectral index, $n_s$, and the tensor to scalar fluctuation ration, $r$, 
are presented in \cite{FHR}. An acceptable value for $n_s$ is possible for 
$|\alpha_1  |<0.1$. The value of $r$ is sensitive to $\alpha_2$ and is 
between one and two orders of magnitude less than the current observational
bound for $|\alpha_2 |>1.$

\subsection{The ``reheat" phase}
Once $\epsilon\approx 1$, the slow-roll conditions are violated and there is a period of 
rapid field oscillation - the ``reheat'' phase in which the scalar fields acquire large 
kinetic energy. We have not been able to find an analytic solution in this phase but a numerical 
study confirms this is the case. 

An example is shown in Fig.(\ref{figure_back}) where it may be seen that after 
about 150 e-folds of inflation the Hubble parameter drops very rapidly before 
rolling to the infra-red fixed point value. As the Hubble parameter drops the 
fields undergo very rapid oscillations (too rapid to show up in the Figure) after 
which they re-enter the slow-roll regime with values in the domain of attraction 
of the IR stable fixed point. During the ``reheat" phase, and all
subsequent evolution,  the kernel, K, remains constant.

\subsection{Infrared fixed point}

After the "reheat" phase the fields enter a second slow-roll phase that is again described by eq(\ref{slowroll}). One may see that this equation has an IR stable fixed point given by 
\beq
x(t\rightarrow\infty)=x_0
\eeq
 This corresponds to the final ratio of the field VEV's given by
 \beq
 {\langle \chi_f\rangle^2\over\langle\phi_f\rangle^2}={\alpha_2\lambda-\alpha_1 \delta\over \alpha_1 \xi-\alpha_2\delta}
 \eeq
A large hierarchy between the ``matter'' sector scale and the Planck scale requires 
that the $\chi$ mass be hierarchically small compared to the Planck scale and 
this in turn requires $\delta\le{\langle\chi_f\rangle^2/\langle\phi_f\rangle^2}$. 
In addition it is desirable that the cosmological constant after inflation be small or zero and this in turn 
requires a fine tuning of the parameters in the potential so that it is (or is close to) a perfect square. 
For this to happen we need $\lambda\le {\langle\chi_f\rangle^4/\langle\phi_f\rangle^4}$. 
Note that these choices are consistent 
with our assumption that $W\sim\xi\chi^4$ and $B\gg A$ during inflation when $\phi$ and $\chi$ are both large.

What happens to the scale factor in the IR?  For static scalar fields the FRW equation is
\be
3M^2\left({\dot{a}\over a}\right)^2=W=\left(\frac{\lambda}{4}+\frac{\xi \mu^4}{4}+\frac{\delta\mu^2}{2}\right)\phi_0^4
\label{1.2a}
\ee
(where 
 $\mu^2\equiv {\langle\chi_f\rangle^2}/{\langle\phi_f\rangle^2}$) and we can define an effective cosmological 
constant $\Lambda_{\rm eff}=(\lambda/4+\xi \mu^4/4+\delta\mu^2/2)\phi_0^2/(\alpha_1 +\alpha_2\mu^2).$  
With the ordering of the couplings discussed above $\Lambda_{eff}\le \xi\chi_f^4/4M_P^2$. If this is non-zero 
there will be a late stage of eternal inflation. To obtain zero cosmological constant 
requires fine tuning of the couplings corresponding to the potential having the form of a perfect square.
 

\subsection{The Dilaton}

 The dilaton effective action
can be derived in analogy to the single scalar case in Section II C
(see IV. B below).
Once the ratio of fields is fixed, the dilaton can readily be identified in the two scalar case 
from the fact that the scale current has the form $K_\mu\propto \partial_\mu \sigma$ and under a 
scale transformation $\sigma\rightarrow\sigma +\epsilon$.  Since the scale current has the form 
$K_\mu=\partial_\mu K$ with K given by eq(\ref{kk2}) we know that $\sigma $ must be some 
function of $K$. In order for scale symmetry to act as a shift symmetry implying
\beq
K={1\over 2}f^2e^{2\sigma/f}
\eeq
with
\beq
f=\sqrt{2K_0}= \sqrt{(1-\alpha_1 )\phi_0^2+(1-\alpha_2)\chi_0^2}
\eeq 
Upon passing to 
the ``Einstein frame,'' the dilaton $\sigma$ appears in the action only in its kinetic term as 
for the single scalar case, eq(\ref{fixed2}).
The dilaton decoupling is due to the exact underlying global Weyl invariance
that is broken only spontaneously via the VEV of K. This will be discussed
in detail elsewhere \cite{FHR2}.


{\section{N-scalar case}

The analysis generalizes readily to the case of N-scalars. Here
 the scale current and its associated kernel are derived and the dilaton identified. 
It is also shown that the IR fixed point structure determines the ratios of all the scalar field VEV's in terms of the couplings entering the potential,  so n hierarchical structure can emerge if the couplings are themselves hierarchical. 

However, the existence of an initial inflationary era needs to be justified 
if there are large couplings between the fields 
as this can prevent a period of slow-roll from occurring. This is of particular relevance 
if we treat the additional scalars as a model for the low-energy ``matter''
sector, for then there is no reason why the couplings should be anomalously small. 
To illustrate this we consider below the case of $3$ scalar fields, $\phi_i$, with 
large self and cross couplings between the two ``matter" fields.

\subsection{ $N$-Scalar Action}
The mathematical generalization to $N$-scalars is straightforward.
Consider a set of $N$ scalar quantum fields $\phi _{i},$ $i=(1,2,...N)$ and
action:
\bea
\label{actionN}
 S &= &\int \sqrt{-g}\left( \underset{i}{\sum }\frac{1}{2}g^{\mu \upsilon
}\partial _{\mu }\phi _{i}\partial _{\nu }\phi _{i}-W\left( \phi _{i}\right)
-\underset{i}{\sum }\frac{\alpha _{i}\phi _{i}^{2}}{12}R\right) 
\nonumber \\
\!\!\!\!\!\!\!\!\!\!\!\!\!\!\!
& & 
\!\!\!\!\!\!\!\!\!\!\!\!\!\!\!\!\!\!\!\!\!\!\!\!\!\!\!\!\!
\eea
The Einstein equation is:
\bea 
\label{aa}
\frac{1}{6} \underset{i}{\sum }\alpha _{i}\phi _{i}^{2}
G_{\alpha \beta }& = & g_{\alpha \beta }W(\phi_{i})
\nonumber \\
& &  
\!\!\!\!\!  \!\!\!\!\! \!\!\!\!\! \!\!\!\!\!   \!\!\!\!\!  \!\!\!\!\! \!\!\!\!\! \!\!\!\!\! 
 \!\!\!\!\!  
+\underset{i}{\sum }\left[\left(1-\frac{\alpha _{i}}{3}\right)\partial
_{\alpha }\phi _{i}\partial _{\beta }\phi _{i}-\left( \frac{1}{2}
-\frac{\alpha _{i}}{3}\right) g_{\alpha\beta } \partial
^{\mu }\phi _{i}\partial _{\mu }\phi _{i} \right.
\nonumber \\
& &  
\!\!\!\!\!  \!\!\!\!\! \!\!\!\!\! \!\!\!\!\!   
\left. +
\left(\frac{\alpha _{i}}{3}\right)\left( g_{\alpha \beta }\phi _{i}D^{2}\phi _{i}-\phi
_{i}D_{\beta }D_{\alpha }\phi _{i}\right)\right]
\eea 
The trace of the Einstein equation becomes:
\bea
\label{traceN}
 \!\!\!\!\!\!\!\!\!\! \!\!\!\!\!\!\!\!\!\!\!\!\!\!\! 
-\frac{1}{6}\left( \underset{i}{\sum }\alpha _{i}\phi _{i}^{2}\right) R 
& = &  4W(\phi) 
\nonumber \\ 
& &  \!\!\!\!\!  \!\!\!\!\! \!\!\!\!\! \!\!\!\!\!\!\!\!\!\!  \!\!\!\!\! \!\!\!\!\! 
+\underset{i}{\sum }\left[(\alpha _{i}-1)\partial ^{\mu }\phi _{i}\partial _{\mu
}\phi _{i}
+\alpha _{i}\phi _{i}D^{2}\phi _{i}\right] 
\eea
The $N$ Klein-Gordon equations are:
\beq
0 = D^2\phi_i +\frac{\delta }{\delta \phi _{i}}W(\phi )+\frac{1}{6}\alpha_i\phi_i R
\eeq
and we can write the sum of the Klein-Gordon equations:
\bea
\label{KGN}
\!\!\!\!\! \!\!\!\!\! 
 -\frac{1}{6}\left( \underset{i}{\sum }\alpha_{i}\phi _{i}^{2}\right) R
& = &
\underset{i}{\sum }\phi _{i}D^{2}\phi _{i}+\phi
_{i}\frac{\delta }{\delta \phi _{i}}W(\phi )
\eea
Combine eqs.(\ref{traceN},\ref{KGN})
to eliminate $R$:
\bea
\label{elimR}
0& =& \underset{i}{\sum }\left[ (\alpha _{i}-1)\partial ^{\mu }\phi _{i}\partial
_{\mu }\phi _{i}+\left( \alpha _{i}-1\right) \phi
_{i}D^{2}\phi _{i}\right]
\nonumber \\
& & \qquad +4W\left( \phi \right) -\phi _{i}\frac{\delta }{\delta
\phi _{i}}W\left( \phi \right) 
\eea
If we assume a scale invariant potential
we have:
\bea
\label{scaleN}
 0 & = & 4W\left( \phi \right)
-\underset{i}{\sum }\phi _{i}\frac{\delta }{\delta \phi _{i}}W\left( \phi \right) 
\eea
We thus see that eqs.(\ref{elimR},\ref{scaleN})
implies a covariantly conserved current:
\bea
\label{KcurrentN}
{K}_{\mu }=\underset{i}{\sum }\left(1- \alpha
_{i}\right) \left( \phi _{i}\partial_{\mu }\phi _{i}\right)
\eea
where   $D_\mu {K}^{\mu }=0$.
The current $ {K}_{\mu }$  arises from a ``Weyl gauge transformation''  and
the $K_\mu$ current has a ``kernel,'' \ie, it can be written as a gradient,
$K_\mu = \partial_\mu K$ where:
\beq
\label{Nconstraint}
K=\frac{1}{2}\underset{i}{\sum }\phi^2_{i}(1-\alpha _{i})
\eeq

\subsection{$N$-Scalar Dilaton}


The  scale symmetry is spontaneously broken by the constraint
of eq.(\ref{Nconstraint}). The fixed value of $K$ has
been generated inertially by the dynamical dilution of thebibliographycharge density, $K_0$.
The value of $K$ is arbitrary and it can be be shifted at no cost in energy
due to overall Weyl invariance.
This implies a dilaton.  We can define the dilaton as:
\beq
\sigma={f\over 2}\log\left({2K\over f^2}\right)
\eeq 
To obtain the dilaton action we perform a local Weyl tranformation 
using the dilaton field itself:
\bea
g_{\mu \nu }(x) &\rightarrow & \exp (-2\sigma(x)/f)g_{\mu \nu }(x)
\nonumber \\
\phi _{i}(x) & \rightarrow & \exp (\sigma(x)/f)\phi _{i}(x)
\eea
Hence, the action $S$ of eq.(\ref{actionN}) becomes $S+\delta S$ with:
\bea
\delta S 
& = & 
\int \sqrt{-g}\left[\frac{1}{f} \underset{i}{\sum }(1-\alpha _{i})  \phi_{i}\partial_{\mu }\phi_{i}
 \left( \partial ^{\mu }\sigma(x)\right)\right.
\nonumber \\
& & \qquad \left.
 +\frac{1}{2f^2}\underset{i}{\sum }(1-\alpha _{i})\phi _{i}^{2}\left(
\partial_{\rho }\sigma (x)\partial^{\rho }\sigma (x)\right) \right]
\nonumber \\
\!\!\!\!\!\!\!\! \!\!\!\!\!\!\!\! 
& = & 
\int \sqrt{-g}\left[\frac{1}{f}{K}_{\mu }
 \left( \partial ^{\mu }\sigma(x)\right)
 + \frac{K}{f^2}\left(
\partial_{\rho }\sigma (x)\partial^{\rho }\sigma (x)\right) \right]
\nonumber \\
\eea
This implies
\beq
f=\sqrt{2K} 
\eeq
is the dilaton decay constant, (for constant $K$).
We can integrate the first term by parts and use
the covariant  $K_\mu$ current divergence, $D_\mu K^\mu =0$,
leaving a decoupled dilaton in the Einstein frame.
Technically, we should include a Lagrange multiplier to enforce
the constraint of eq.(\ref{Nconstraint}) on the $\phi_i$.

\subsection{Slow-roll}
\label{slowrollN}

 The evolution equations take the form:
\begin{eqnarray}
\!\!\!\!\!  \!\!\!\!\! \!\!\!\!\! \!\!\!\!\! \!\!\!\!\!
\left( \begin{array}{cccc}
1+\frac{\alpha_1^2\phi_1^2}{6M^2} &\frac{\alpha_1\alpha_2\phi_1\phi_2}{6M^2}& \cdots& \frac{\alpha_1\alpha_N\phi_1\phi_N}{6M^2}\\
\frac{\alpha_1\alpha_2\phi_1\phi_2}{6M^2} &1+\frac{\alpha^2_2\phi_2^2}{6M^2}& \cdots& \frac{\alpha_2\alpha_N\phi_2\phi_N}{6M^2}\\
\cdots&\cdots& \cdots& \cdots\\
\frac{\alpha_1\alpha_N\phi_1\phi_N}{6M^2} &\frac{\alpha_2\alpha_N\phi_2\phi_N}{6M^2}&\cdots& 1+\frac{\alpha_N^2\phi_N^2}{6M^2}  \end{array} \right)\left( \begin{array}{c}
3H{\dot \phi}_1\\
3H{\dot \phi}_2\\
\cdots\\
3H{\dot \phi}_N\end{array} \right)\nonumber 
\hspace{-1.5cm}\\
=-\left( \begin{array}{c}
\frac{4\alpha_1\phi_1}{6M^2}W+W_{\phi_1}\\
\frac{4\alpha_2\phi_2}{6M^2}W+W_{\phi_2}\\
\cdots\\
\frac{4\alpha_N\phi_N}{6M^2}W+W_{\phi_N}\\
\end{array} \right)
\end{eqnarray}
As before we assume that $U\equiv \lambda_N\phi^4_N$ dominates. 
We then have:
\begin{eqnarray}
\left( \begin{array}{c}
\frac{4\alpha_1\phi_1}{6M^2}W+W_{\phi_1}\\
\frac{4\alpha_2\phi_2}{6M^2}W+W_{\phi_2}\\
\cdots\\
\frac{4\alpha_N\phi_N}{6M^2}W+W_{\phi_N}\\
\end{array} \right)=
\frac{4U}{6M^2}\left( \begin{array}{c}
\alpha_1\phi_1\\
\alpha_2\phi_2\\
\cdots\\
-\frac{\sum_i^{N-1}\alpha_i\phi_i^2}{\phi_N}\\
\end{array} \right)
\end{eqnarray}
We can now solve this system  to get:
\begin{eqnarray}
\!\!\!\!\!\!\!\!\!\! 
-3H\left( \begin{array}{c}
{\dot \phi}_1\\
{\dot \phi}_2\\
\cdots\\
{\dot \phi}_N\end{array} \right)=\frac{4U}{\sum_i^N\alpha_i(1-\alpha_i)\phi^2_i}\left( \begin{array}{c}
-\alpha_1(1-\alpha_N)\phi_1\\
-\alpha_2(1-\alpha_N)\phi_2\\
\cdots\\
\frac{\sum_i^{N-1}\alpha_i(1-\alpha_i)\phi_i^2}{\phi_N}\\
\end{array} \right)
\nonumber \\
\end{eqnarray}
We now define $X_i=\alpha_i\phi^2_i$ to get:
\begin{eqnarray}
\!\!\!\!\!\!\!\!\!\!
-\frac{3}{2}H\left( \begin{array}{c}
{\dot X}_1\\
{\dot X}_2\\
\cdots\\
{\dot X}_N\end{array} \right)
& = & \frac{4U}{\sum_i^N(1-\alpha_i)X_i}
\times
\nonumber \\
& & 
\!\!\!\!\!\!\!\!\!\!
\left( \begin{array}{c}
-\alpha_1(1-\alpha_N)X_1\\
-\alpha_2(1-\alpha_N)X_2\\
\cdots\\
\alpha_N\sum_i^{N-1}(1-\alpha_i)X_i\\
\end{array} \right)
\end{eqnarray}
If we now change variables to $\ln a$ and use the FRW equation:
\begin{eqnarray}
3H^2 =\frac{U}{M^2}
\end{eqnarray}
we get:
\begin{eqnarray}
\!\!\!\!\!
\left( \begin{array}{c}
{X}'_1\\
{X}'_2\\
\cdots\\
{X}'_N\end{array} \right)=\frac{4}{3}\frac{\sum_i^NX_i}{\sum_i^N(1-\alpha_i)X_i}\left( \begin{array}{c}
-\alpha_1(1-\alpha_N)X_1\\
-\alpha_2(1-\alpha_N)X_2\\
\cdots\\
\alpha_N\sum_i^{N-1}(1-\alpha_i)X_i\\
\end{array} \right)
\nonumber \\
\!\!\!\!\!\!\!\!\!\! \!\!\!\!\!\!\!\!\!\! \!\!\!\!\!\!\!\!\!\!
\end{eqnarray}
If we now take the $X_N\gg X_i$ (with $i=1,\cdots,N-1$) we get:
\begin{eqnarray}
\left( \begin{array}{c}
{X}'_1\\
{X}'_2\\
\cdots\\
{X}'_N\end{array} \right)=\frac{4}{3}\left( \begin{array}{c}
-\alpha_1X_1\\
-\alpha_2X_2\\
\cdots\\
\frac{\alpha_N}{1-\alpha_N}\sum_i^{N-1}(1-\alpha_i)X_i\\
\end{array} \right)
\nonumber \\
\end{eqnarray}
We can solve with $\nu_i=-\frac{4}{3}\alpha_i$ and $\gamma_i=\frac{\alpha_N(1-\alpha_i}{\alpha_i(1-\alpha_N)}$:
\begin{eqnarray}
X_i&=& X^{(0)}_ie^{\nu_i \ln a} \ \ \ \mbox{$i=1,\cdots,N-1$} \nonumber \\
X_N&=& C+\sum_{i=1}^N\gamma_i X^{(0)}_ie^{\nu_i\ln a}
\end{eqnarray}

\subsection{Fixed point structure}

The fixed points are found  solving the $N$ equations:
\begin{eqnarray}
\frac{4\alpha_i\phi}{6M^2}W+W_{\phi_i}=0
\end{eqnarray}
We can rewrite this:
\begin{eqnarray}
4\alpha_i\phi_i\sum_{jk}\phi^2_jW_{jk}\phi^2_k-4\sum_{j}\alpha_j\phi^2_j\sum_k\phi_iW_{ik}\phi^2_k=0
\end{eqnarray}
We divide out $\alpha_i\phi_i$ and define a set of N matrices (labelled by $i$):
\begin{eqnarray}
{\cal A}^{(i)}_{jk}=W_{jk}-\frac{\alpha_j}{\alpha_i}W_{ik}
\end{eqnarray}
We then have that the N quadratic forms satisfy:
\begin{eqnarray}
\sum_{jk}\phi^2_j{\cal A}^{(i)}_{jk}\phi^2_k=0
\end{eqnarray}
If this is to be possible then we must have ${\rm Det}[{\cal A}]=0$. But this is
trivially so. If we pick the $i$th matrix, it will have that its $i$th line will be:
\begin{eqnarray}
{\cal A}^{(i)}_{ik}=W_{ik}-\frac{\alpha_i}{\alpha_i}W_{ik}=0
\end{eqnarray}
which means that its rank is less than or equal than $N-1$. If all the $\alpha_i$ are
different, and if we assume $W_{ik}$ is non-singular, we have that the rank is $N-1$ 
and the solution will be a line in $\phi^2_i$ space with one free parameter, 
the overall scale. Interestingly, if some of the $\alpha_i$ are degenerate, 
then the subspace will have a higher dimensionality.

\subsection{ Slow-roll in a $3$-scalar scheme}

The fixed point structure proves to be important in the slow-roll regime for
the case that more than one coupling is significant in the scalar potential 
during slow-roll. We illustrate this presently in a particular 3-scalar 
example. Consider the case that the significant couplings
during slow roll involve only two ``matter'' fields,
$\phi_2$ and $\phi_3$. In this case the  potential is
dominantly of the form $W=U+Y+T$ where: 
\beq
U = a{\phi_2 ^4},\;\;T = b{\phi_3 ^4},\;\;V = c{\phi_2 ^2}{\phi_3 ^2}.
\eeq
In writing the slow-roll equations it is convenient to define new fields:
\beq
X =  - \alpha_1 {\phi_1 ^2},\;\;Y =  -\alpha_2 {\phi_2 ^2},\;\;Z =  - \alpha_3 {\phi_3 ^2}.
\eeq
Here $\phi_2$ and $\phi_3$ are the ``matter'' fields and we allow $a,\;b$ and $c$ to be $O(1)$.
Then the evolution  equations in the slow-roll region have the form: 
\bea
\begin{array}{l}
\left( {\begin{array}{*{20}{c}}
{X'}\\
{Y'}\\
{Z'}
\end{array}} \right) =  - \frac{4}{3}.\frac{{X + Y + Z}}{{\beta_1X + \beta_2Y + \beta_3Z}}.\frac{1}{{(U + T + V)}}\times
\nonumber \\
\quad \quad \quad \quad\\
 \left( {\begin{array}{*{20}{c}}
{ - \alpha_1 X\left\{ {\beta_2\;(U + V) + \beta_3\;(T + V)} \right\}}\\
{\alpha_2 \left\{ {\beta_1X(\,U + {V\over2}) + \beta_3\left[ {Z\,(U + {V\over 2}) - \,Y(T + {V\over 2})} \right]} \right\}}\\
{\alpha_3 \left\{ {\beta_1X\,(T + {V\over 2}) + \beta_2\left[ {Y\,(T + {V\over 2}) - Z\,(U + {V\over 2})} \right]} \right\}}
\end{array}} \right)
\end{array}
\nonumber \\
\label{RGE3}
\eea
where $\beta_i=1-\alpha_i$.

The problem is that, even if $\alpha_i$ are very small,  
the large couplings $a,\;b$ and $c$ cause the fields $Y$ and $Z$ to roll 
quickly and violate the slow-roll conditions used to derive the evolution equations.  
In the small  $\alpha_1$  regime  we see from 
eq.(\ref{RGE3}) that the dominant terms  are proportional to,   
$ \pm (Z\,(U + V/2) - \,Y(T + V/2))$,
  respectively, with positive coefficients. These terms have an IR stable fixed point with:
\beq
Z\,(U + V/2) = \,Y(T + V/2),\quad i.e.\;\frac{{{\phi_2 ^2}}}{{{\phi_3 ^2}}} = 
\frac{{2b\alpha_2  - c\alpha_3 }}{{2a\alpha_3  - c\alpha_2 }}
\eeq
At this fixed point the evolution equation becomes: 
\bea
\begin{array}{l}
\left( {\begin{array}{*{20}{c}}
{X'}\\
{Y'}\\
{Z'}
\end{array}} \right) =  - \frac{4}{3}.\frac{{X + Y + Z}}{{\beta_1X + \beta_2Y + \beta_3Z}}.\frac{1}{{(U + T + V)}}
\nonumber \\
\quad \quad \quad \quad
 \left( {\begin{array}{*{20}{c}}
{ - \alpha_1 X\left\{ {\beta_2\;(U + V) + \beta_3\;(T + V)} \right\}}\\
{\alpha_2 \beta_1X(\,U + {V\over2})  }\\
{\alpha_3  \beta_1X\,(T + {V\over 2}) }
\end{array}} \right)
\end{array}
\nonumber \\
\eea

As in the two scalar case all derivatives are proportional to X. Since $X'$ is proportional to $\alpha_1$, 
if $\alpha_1$ is small the slow-roll constraints can indeed be satisfied. Also the evolution of 
Y and Z is much faster than X in the  $\alpha_1\ll \alpha_{2,3}$ regime, so the inflationary 
era in the three scalar case will be similar to that in the two scalar case.

\section{Quantum Effects and the $K_\mu$ Current}

{We now consider the quantum effects. We first give
a formal derivation of the conventional anomalies of the $K_\mu$ current,  and
show how this is realized in a Coleman-Weinberg-Jackiw effective action. We  then discuss how
Weyl invariance can be maintained in the renormalized theory.  This implies that renormalized quantities 
satisfy renormalization
group equations in which they run in Weyl invariant combinations of fields, such as
the ratios of scalar fields. The trace
anomaly is then absent and the $K_\mu$ current is identically conserved.}

\subsection{Weyl Invariance and Effective Action}

Scale symmetry of a theory is normally
considered to be broken by quantum loops.
However, this happens because at some stage in the 
renormalization procedure,  we
introduce explicit ``external'' mass scales into the theory by hand. These are mass scales
that are not part of the defining action of the theory, and 
they lead to  non-conservation of the scale current.

{The renormalization
procedure, however, can be made  scale invariant if we specify these
quantities,  not by introducing external mass scales,
but rather by using the VEV's of scalar fields that spontaneously break the scale symmetry
but are part of the action itself. 
In this case, all logarithmic corrections arising in loops will have as their arguments 
scale invariant ratios of the internal field VEV's. At the formal
level, which we develop presently,  the choice of dependencies of renormalized quantities
appears arbitrary. However, calculations can be
performed in which this arbitrariness is removed, and we will discuss
this elsewhere \cite{FHR2}.}

 We can see the usual ``external mass parameter'' renormalization in
the famous paper of Coleman and Weinberg \cite{CW}. 
Starting with the classical $\lambda\phi^4/4$ theory, in their eq.(3.7) to renormalize
$\lambda$ at one-loop level,
they introduce a mass scale $M$.  Once one injects $M$ into the theory, 
one has broken scale symmetry.
The one-loop effective potential then takes the form: 
\beq
\label{V1}
V(\phi)=\frac{\beta_\lambda}{4} \phi^4 \ln(\phi/M).
\eeq
where $\beta_\lambda $ 
is the one-loop  approximation, ($\cal{O}(\hbar)$),
to the $\beta$--function, $\beta_\lambda = {d \lambda(\mu)}/{d\ln \mu}$.

The heart of our proposal is to replace $M$ by the VEV of another dynamical field, 
\eg, $\chi$, that is part of the action of our theory:
 \beq
 \label{V2}
V =\frac{\beta_\lambda}{4} \phi^4 \ln(\phi/\chi).
\eeq
We see that the Weyl symmetry, $\phi\rightarrow e^\epsilon\phi$, 
$\chi\rightarrow e^\epsilon\chi$, is now intact.

The manifestation of this 
can be seen in the trace of the improved stress tensor \cite{CCJ}.\footnote{ Technically,
the improved stress tensor is defined only for $\alpha=1$,
and in the flat space limit, but its anomaly parallels that of the $K_\mu$ current;
the $K_\mu$ current is the more relevant scale current for $\alpha\neq 1$ theories.}
In a single scalar theory, the trace anomaly is the divergence
of the scale current $S_\mu$ and, using eq.(\ref{V1}),  is given by \cite{Hill}:
\beq
\label{traceanom1}
\partial_\mu S^{\mu} =T_\mu^\mu=4V(\phi)-\phi\frac{\partial}{\partial\phi}V(\phi) =-\frac{\beta_\lambda}{4} \phi^4
\eeq
We see, as usual, that the trace anomaly is
directly associated with the $\beta$-function
of the coupling constant $\lambda$, and it exists on the {\em rhs}
of eq.(\ref{traceanom1}) because we have introduced the
explicit scale breaking into the theory
by hand via $M$. On the other hand, with two scalars we have:
\beq
\label{traceanom2}
\partial_\mu S^{\mu} =T_\mu^\mu=4V-\phi\frac{\partial}{\partial\phi}V-\chi\frac{\partial}{\partial\chi}V
=0
\eeq
and this is vanishing with eq.(\ref{V2}).  In effect, the trace anomaly has been transferred
onto the {\em lhs} of the divergence equation, and the overall scale current conservation is maintained.
We will see that this applies to the Weyl current $K_\mu$ as well.

Of course, there's nothing wrong with the Coleman-Weinberg
procedure, if one is only treating the effective
potential as a subsector of the larger theory.  
That is, we are simply deferring the question 
of what is the true origin of $M$ in the larger theory?
If, however, scale symmetry is to be maintained as an exact invariance
of the world, then  $M$ must be replaced by an internal mass scale that is part of action, 
\ie, $M$ must then be the VEV
 a field appearing in the extended action, such as $\chi$. 
If $M$ is replaced by  a dynamical field
in our theory,  we will still have renormalization
group evolution, but the resulting physics can now
depend only upon ratios of dynamical VEV's, and the running of couplings
is given in terms of these ratios.  

In fact, this is something we
do in practice. All mass scales we measure in the laboratory
are referred to other mass scales. Even derived
scales, such as $\Lambda_{QCD}$ can be viewed as arising from a specification of
$\alpha_{QCD}$ at some higher energy scale, such as a grand-unification scale, or
the Planck mass, $M_{Planck}$. With the boundary condition, specifying $\alpha_{QCD}(M_{Planck})$
then $\Lambda_{QCD}$ is computed from the solution to the renormalization group equation. 
We obtain
$\Lambda_{QCD} =c M_{Planck}$, where $c$ is an exponentially small coefficient
(at one loop $c \sim \exp(-2\pi/|b_0|\alpha_{QCD}(M_{Planck}))$.
The question is then whether the fundamental reference scale, usually
taken to be $M_{Planck}$, is an external input scale (such as the string constant),
or the dynamical VEV of a field, such as $\chi$?  In the latter case, we can in principle 
maintain an overall Weyl symmetry,
and derived mass scales, such as $\Lambda_{QCD}$ become Weyl covariant:
$\chi\rightarrow e^\epsilon \chi$, $\Lambda_{QCD}\rightarrow e^\epsilon \Lambda_{QCD}$.

\subsection{Conventional Anomalies of the $K_\mu$ Current}

{Let us first formulate the  anomalies of the $K_\mu$ current
in the conventional renormalization
framework that introduces an external mass scale $M$,} 
in a theory with fields $\phi,\; g_{\mu \nu },\; ...$  The Weyl transformation is:
\beq
\phi \rightarrow e^{\epsilon }\phi ,\qquad  g_{\mu \nu }\rightarrow e^{-2\epsilon }g_{\mu \nu } ,\qquad ...
\eeq
The contravariant metric must then transform as , $g^{\mu \nu }\rightarrow
e^{2\epsilon }g^{\mu \nu }$.
Here, if $\epsilon (x)$ is a function of spacetime the transformation is
local; if $\epsilon $ is a constant in 
spacetime the transformation is global.

It is useful to define a differential operator that acts upon fields:
\beq
\label{Wtrans1}
\delta _{W}\phi = \phi\;\delta\epsilon, \qquad \delta _{W}g_{\mu \nu }=-2g_{\mu \nu }\;\delta\epsilon.
\eeq 
$\delta _{W}$ acts distributively, and, $\delta _{W}g^{\mu \nu }=+2g^{\mu \nu }\delta\epsilon $, 
$\delta _{W}\left( \phi ^{-1}\right) =-\phi ^{-1}\delta \epsilon $,
and $\delta _{W}\left( \ln \phi \right) =\phi
^{-1}\delta _{W}\left( \phi \right) =\delta\epsilon .$ In general, a field $\Phi $ 
of ``mass dimension $D$''  transforms  covariantly  as $\Phi \rightarrow
e^{D\epsilon }\Phi $ or $\delta _{W}\Phi =D\Phi \delta\epsilon .$

 Any locally Weyl  invariant functional of fields $Q(\phi, g_{\mu\nu}..)$ satisfies: 
\beq
\delta_{W} Q=0
\eeq
We typically seek an effective Coleman-Weinberg-Jackiw action as a functional
of classical fields for the study
of inflation and spontaneous scale generation.

For the single scalar field $\phi$,
consider the effective action, constructed by adding sources to the fields, performing
a Legendre transformation to the classical background fields, and integrating out quantum fluctuations
\cite{CW,Jackiw10}.
The result for a single scalar field theory is  a functional of local classical 
background fields $\phi(x)$
and $g_{\mu\nu}(x)$:
\bea
\label{CWone}
S& =&\int \sqrt{-g}\left( \frac{1}{2}g^{\mu \upsilon }\partial _{\mu }\phi
\partial _{\nu }\phi 
 -\frac{\lambda(\phi,g)}{4}\phi^4 
 -\frac{\alpha(\phi,g)}{12}  \phi^{2}R\right)
 \nonumber \\
\eea
It is important to maintain locality in the Lagrangian,
since general covariance is a local symmetry, and therefore requires that
effective coupling constants be local functions
of the fields.

Computing $\delta_W S$ we obtain the difference between the Einstein trace
equation and the Klein-Gordon equations that yields the conservation
law for $K_\mu$.  This calculation is simplified by noting the local Weyl invariants
satisfy:
\bea
&& \delta_W \int \sqrt{-g}\left( \frac{1}{2}g^{\mu \upsilon }\partial _{\mu }\phi
\partial _{\nu }\phi  -\frac{1}{12}  \phi^{2}R\right)=0
 \nonumber \\
&& \delta_W \int \sqrt{-g}\phi^4=0
\eea
Hence:
\bea
\label{CWKdiv1}
\delta_W S \! &=& -\int \sqrt{-g}\delta\epsilon (
 D^\mu K_\mu 
+ \!\frac{1}{4}(\delta_W \lambda) \phi^4+ \! \frac{1}{12}(\delta_W\alpha) \phi^2R )
\nonumber \\
& = & 0
\eea
where we integrate terms with 
$\partial_\mu (\delta \epsilon)$, by parts and discard surface terms.
$K_\mu$ is given by the usual expression, but now
contains the field dependent $\alpha(\phi,g_{\alpha\beta})$ :
\beq
K_\mu= \half \partial_\mu(1-\alpha(\phi,g_{\alpha\beta}))\phi^2
\eeq
Eq.(\ref{CWKdiv1}) defines the anomaly of the current:
\beq
\label{CWKdiv11}
D^\mu K_\mu 
=- \frac{1}{4}(\delta_W \lambda) \phi^4- \frac{1}{12}(\delta_W\alpha) \phi^2R)
\eeq

Consider the theory in a limit where
we ignore all but internal $\phi$ loops.
If we renormalize
the effective action, introducing  an external mass scale, $M$, then the
$\beta$-functions are:
\bea
\label{RGmath1}
\phi\frac{\partial \lambda}{\partial\phi}& = & \beta_\lambda \;\;\left(=\frac{9\lambda^2}{8\pi^2}\right)
\nonumber \\
\phi\frac{\partial\alpha}{\partial\phi} & =  & \beta_\alpha=(\alpha-1)\gamma_\alpha
\qquad
\left(\gamma_\alpha  =  \frac{3\lambda}{8\pi^2}\right)
\eea
where in brackets we quote the 1-loop computed values that follow from the $\phi$ loops in
this theory.

Renormalizing with an external mass scale $M$ implies the constraint:
\beq
\label{RGmath2}
0 =  \phi\frac{\partial \lambda}{\partial\phi}+M\frac{\partial \lambda}{\partial M};
\qquad
0 =  \phi\frac{\partial\alpha}{\partial\phi}+M\frac{\partial \alpha}{\partial M}
\eeq
The ${\partial }/{\partial M}$ terms in the above equations 
are not due to the loop calculations, but rather, are external conditions we impose upon the couplings.
That is, eq.(\ref{RGmath2}) defines the functional dependence of
the counterterms in the theory upon the external mass parameter $M$.

Note that
the RG equation for $\alpha$ is $\propto (\alpha-1)$, which is why
we introduce the factor $\gamma_\alpha$
into its $\beta$-function definition.
We can write $\phi\partial\alpha'/\partial\phi  =  \alpha'\gamma_\alpha$
where, $\alpha' = \alpha-1$, and this leads, for
approximately constant $\gamma_\alpha$, to the solution eq.(\ref{RGrun}) below.
The solutions to the RG equations in the approximation
of a fairly constant or small $\lambda$, \ie, small $\beta_\lambda$,  are, 
\beq
\label{RGrun}
\lambda(\phi) =\beta_\lambda\ln\left(\frac{c\phi}{M}\right)\qquad
\alpha(\phi) =1+(\alpha_0-1)\left(\frac{\phi}{M}\right)^{\gamma_\alpha}
\eeq
where the constants $c $ and $\alpha_0$  define the RG trajectories of the
running couplings, $\lambda$ and $\alpha$.

Eq.(\ref{CWKdiv11}) with eq.(\ref{RGrun}) then implies the form
of the $K_\mu$ anomalies:
\bea
\label{CWKdiv2}
D^\mu K_\mu & = & -\frac{1}{4}\beta_\lambda \phi^4 - \frac{1}{12}\beta_\alpha \phi^2R
\eea
The non-conservation of the $K_\mu$ current arises because the external mass parameter,
$M$, breaks the Weyl scale symmetry.

Armed with the solutions of eqs.(\ref{RGrun}) we then have the 
effective action, where Weyl symmetry is broken by the effect of $M$:
\bea
\label{CWtwo}
\!\!\!\!\! \!\!\!\!\!   S & =&\int \sqrt{-g}\left( \frac{1}{2}g^{\mu \upsilon }\partial _{\mu }\phi
\partial _{\nu }\phi 
 -\frac{1}{4}\beta_\lambda\ln\left(\frac{c\phi}{M}\right)\phi^4 \right.
\nonumber \\
& & \;\;\;\;\; \left.
 -\frac{1}{12} \left(1+(\alpha_0-1)\left(\frac{\phi}{M}\right)^{\gamma_\alpha} \right)   \phi^{2}R\right)
\eea
We  cannot have our program of a stable, dynamically generated Planck mass
without maintaining the Weyl symmetry, and we must therefore eliminate
the explicit $M$ dependence and, hence, the anomalies in the $K_\mu$ current.

\subsection{\bf Maintaining Exact Weyl Scale Symmetry in Renormalized Quantum Theory}\label{Sec:exact}

\subsubsection{\bf The Single Scalar Theory}

To preserve the Weyl invariance, we need to eliminate the anomaly,
which requires replacing the constraint eq.(\ref{RGmath2}) that introduces
the external mass scale $M$. 
From eq.(\ref{CWKdiv11}) we see that we can maintain the Weyl invariance of eq.(\ref{Wtrans1})
in the renormalized theory provided the running coupling constants are  Weyl invariant:
\bea
\label{Weylinv}
\delta_W \lambda & = & 0;\;\; \qquad 
\delta_W \alpha  =  0 \eea
Eqs.(\ref{Weylinv}) are thus a new constraint 
that replaces eq.(\ref{RGmath2}).  
Hence, together with eq.(\ref{RGmath1}), imposing eq.(\ref{Weylinv})
 we see from eq.(\ref{CWKdiv1}) that: $D^\mu K_\mu  = 0 $.

This is an almost obvious result: 
\em the coupling constants must be
local functions of Weyl invariants in order to maintain the Weyl symmetry.}
However, 
just as the ${\partial }/{\partial M}$ terms in eq.(\ref{RGmath2})
are not due to the loop calculations, and are really part of
the UV completion of the theory, neither do
the  dependencies upon various compensating fields implicit in eq.(\ref{Weylinv})
necessarily arise from the loops alone.  These are external 
conditions  that presumably come
from the UV completion.

Logically, this procedure is analogous to having a theory in which we have a chiral anomaly
that violates a given axial current which we may want to gauge. This is
usually done explicitly by judicious
choice of fermion representations in the theory.  However, it can also be
done by constructing a Wess-Zumino-Witten term that generates the anomaly through
bosonic fields and can be used to cancel the fermionic chiral anomaly.  
For example, the Wess-Zumino-Witten term for the original Weinberg model
of a single lepton pair $(\nu , e)$, 
can be written in terms of the $0^-$ and $1^-$ mesons 
of QCD, and the $W$, $Z$ and $\gamma$. Including this term into
the original Weinberg model gives the an anomaly free description
for first generation lepton $(\nu, e)$ and the visible states of low energy QCD
(and correctly describes $B+L$ violation, see \cite{HHH}). Of course, this
represents the effects of the underlying confined $(u,d)$ quarks.  
In our present situation we do not know what the underlying Weyl invariant
UV complete theory of gravity and scalars is, but we can imitate the
WZW term by demanding an overall Weyl invariant constraint
that maintains the renormalization group (the $\phi$ loops).

The solutions to the constraint eqs.(\ref{Weylinv}) 
are coupling constants that are functions of
 Weyl invariants.  These  clearly must be Lorentz scalars, 
and also invariant under general coordinate transformations (diffeomorphisms). 
In the single scalar theory, 
we only have  at our disposal the Weyl invariant
objects, 
 $\phi^2 g_{\mu\nu}$, and $\phi^{-2} g^{\mu\nu}$, which
are obviously not scalars.
The quantity $\sqrt{g}\phi^4$ is Weyl invariant, but is a scalar density and not 
diffeomorphism invariant.
This leaves the Ricci scalar, $R(\phi^2 g_{\mu\nu})$, expressed
as a function of the invariant combination $\tilde{g}_{\mu\nu}$
where $\tilde{g}_{\mu\nu}=\phi^2 g_{\mu\nu}$ (and
$\tilde{g}^{\mu\nu}=\phi^{-2} g^{\mu\nu}$):
\beq
R(\phi^2 g)=\phi^{-2}R( g) +6\phi^{-3} g^{\mu\nu}D_\mu\partial_\nu\phi
\eeq
Therefore, we can consider  the arguments
of the logs to be a general functions  $F_i[R(\phi^2 g)]$.
The coupling constants become:
\bea 
\label{RGrun2}
\lambda(\phi) & = & 
\half\beta_\lambda\ln\left(F_\lambda [R(\phi^2 g)]\right)
\nonumber \\
\alpha & = & 1+(\alpha_0-1)\left( F_\alpha [R(\phi^2 g)]\right)^{\gamma_\alpha/2}
\eea 
For example, we might choose:
\beq
\label{choice}
F_i=\frac{c_i\phi^2}{R( g) +\frac{6}{\phi} g^{\mu\nu}D_\mu\partial_\nu\phi + c'_i\phi^2}
\eeq
With the solutions of eqs.(\ref{RGrun}) we have the Weyl invariant Coleman-Weinberg
effective action:
\bea
\label{CWtwo}
S& =&\int \sqrt{-g}\left( \frac{1}{2}g^{\mu \upsilon }\partial _{\mu }\phi
\partial _{\nu }\phi 
 -\frac{1}{4}\beta_\lambda\ln\left(F_\lambda [R(\phi^2 g),\phi^2]\right)\phi^4 \right.
\nonumber \\
& & \!\!\!\!\!    \!\!\!\!\!   \!\!\!\!\!           \left.
 -\frac{1}{12}\left(1+(\alpha_0-1)\left(F_\alpha [R(\phi^2 g),\phi^2] \right)^{\gamma_\alpha/2}\right) \phi^{2}R\right)
\eea
The renormalization group equations 
eq.(\ref{RGmath1}) are now modified:
\bea
\label{RGmath3}
F_\lambda \frac{\partial \lambda}{\partial F_\lambda }& = & \beta_\lambda 
\nonumber \\
F_\alpha \frac{\partial\alpha}{\partial F_\alpha } & =  & (\alpha-1)\gamma_\alpha
\eea

{In writing eq.(\ref{RGmath3}) we
have solved the constraint of eq.(\ref{Weylinv}). Since this is a constraint, 
it only dictates that the functional form of
the $F_i$, be Weyl invariant. In lieu of an exact calculation
is at this stage, the $F_i$ arbitrary. However, such a calculation
of the Coleman-Weinberg potential can be done
(in a simple locally Weyl invariant two scalar theory)
and it yields a specific functional form, as will be presented elsewhere \cite{FHR2}.}

We can specify $F_i$ if we match onto the calculated $\beta$-functions from $\phi$ loops
For example, our choice in eq.(\ref{choice}) will be
consistent with the computed $\beta$-functions of eq.(\ref{RGmath1})
from $\phi$ loops,
(but not necessarily with calculated functions associated with graviton loops).
It is interesting to note that while $\beta_\lambda$ of eq.(\ref{RGmath1})
produces a Landau pole in the running of $\lambda$ with large $\phi$,
the choice of nonzero $c'_\lambda$ implies that asymptotically $\lambda(\phi)$ approaches
a constant, $\lambda(c/c')$.\footnote{ There is a characteristic difference between
RG running in field VEV's and running in momentum space. E.g., 
the top quark, \etc, never decouples if the Higgs VEV runs into the IR.  
RG running for deep scattering processes in momentum will be standard and remains
sensitive to the Landau pole as usual.}

\subsubsection{\bf The Two Scalar Theory}

In the case of the two  scalar scheme,
defined by eqs.(\ref{pot2},\ref{action2}), we 
have 
the five couplings, $(\lambda, \xi, \delta, \alpha_1,\alpha_2)$
and will have 
RG equations for running in $\phi$ or $\chi$.
For the sake of discussion we will presently assume that the field VEV's
$\phi$ and $\chi$ are large compared to curvature $R$.
If we consider a typical coupling constant $\lambda$ we therefore
have the scale invariant constraint:
\bea
\label{Weylinv3}
\frac{\delta_W \lambda}{\delta\epsilon}& = &
 \phi\frac{\partial \lambda}{\partial\phi} + \chi\frac{\partial\lambda}{\partial\chi} =0.
 \eea
We reinterpret
the usual RG equations in terms of $\lambda(F)$
with running in a Weyl invariant function
of $\phi$ and $\chi$, such as an  arbitrary
function of the ratio, $F_\lambda=F(\phi/\chi)$, for example,
$F= \phi/\chi$.
The renormalization group $\beta$-function is now: 
\beq
\label{RG31}
\beta_{\lambda } =  F\frac{\partial \lambda}{\partial F}
\eeq
{ Hence, we can maintain the Weyl symmetry
while having $\beta$-functions that now describe the running of couplings 
in Weyl invariants.  Elsewhere we will demonstrate how to obtain this
result by a direct calculation of the Coleman-Weinberg
effective potential while maintaining a local Weyl symmetry \cite{FHR2}.}

\subsubsection{\bf Relation to other scale invariant schemes}

There have been several proposals for maintaining
Weyl invariance that focus on the regularization schemes \eg, 
see \cite{Percacci,Englert:1976ep,Deser:1970hs,Shaposhnikov:2008xi,ShapoArm, Ghil2, Ghil}.

\vspace{0.5cm}

\noindent {\it (i) Dimensional regularization.} Extensively studied is the case of dimensional 
regularization in which the external mass scale, $\mu$ is replaced by a combination of 
fields, $\mu(\phi,\chi)$. In this approach the Coleman Weinberg formula for the 1-loop 
correction scalar potential:
\beq
-i\int {{d^4}p\;Tr\ln \left[ {{p^2} - {V}\left( {\phi ,\chi } \right) + i\varepsilon } \right]} 
\eeq
is continued to d-dimensions. This gives: 
\beq
V(\phi ,\chi ) = \mu {(\phi ,\chi )^{4 - d}}{V_0}(\phi ,\chi )
\eeq
where ${V_0}(\phi ,\chi )$ is the potential in 4D.  The first factor gives additional 
corrections to  $V$ that, due to the divergent   structure of the integral in 4D, 
give finite contributions to the scalar potential (see \cite{Ghil2, Ghil}). Weyl 
invariance is maintained by choosing  $\mu$ to be a function of $\phi$   and  $\chi$  of 
scaling dimension 1. 

For the very simple choice $\mu  \propto \phi $  the resulting corrections 
are of the form ${\chi ^6}/{\phi ^2} + ...$  and the theory 
must be viewed as an effective field theory valid for   
${\chi ^2}/{\phi ^2}\ll 1.$
Arbitrariness obviously enters here in the choice of $\mu(\phi,\chi)$,
and  will affect the $\beta$-functions as we have discussed above.

\vspace{0.5cm}

\noindent {\it (ii)``Renormalized" perturbation theory.}
In the case of  renormalized perturbation theory the Feynman rules are derived from the 
Lagrangian computed in terms of the physical parameters of the theory. In this case the 
potential will have a dependence on the scale $M$  at which the couplings are determined. 
Writing  $M$ as a function of  $\phi$   and   $\chi$ of scaling dimension 1, Weyl invariance 
can be maintained. However the field dependence of $M=M(\phi,\chi)$   will, as in the case of 
dimensional regularization,  give additional contributions to $M^2$  that  give rise to 
non-renormalisable and arbitrary corrections of the form found in dimensional regularization.

\vspace{0.5cm}

\noindent {\it (iii)``Bare" perturbation theory.}
An alternative possibility is  bare perturbation theory in which the 
Feynman rules are based on the bare Lagrangian.  In this case the bare 
potential has no dependence on the scale  $M$ and so there are no new contributions 
to the potential of the form discussed above. Weyl invariance can be maintained by 
identifying the cut-off scale, $M$, in the loop calculations with a function of the 
fields of scaling dimension 1 and is equivalent to the procedure proposed in Section \ref{Sec:exact}.

\subsection{An ansatz for a quantum corrected theory}

What might be the physical
effects that arise  from  Weyl invariant renormalization? In the following we initially
consider a general form, $F(x)$ for the argument of the log and then specialise the
case where $F=x$. We shall see that this will lead to modifications during inflation to elliptic
path in ($\phi$,$\chi$)  that we described above.


The one-loop CW action (neglecting terms in $\delta$)
 can then take the form of eq.(\ref{action2})
with the potential:
\bea
\label{cwpot0}
W(\phi,\chi)& \simeq & \frac{\lambda\phi^4}{4}+ \frac{\beta_\xi}{4} \chi^4\ln(cF(\phi/\chi))
\nonumber \\
&=& \frac{\lambda\phi^4}{4}\left[1+ \frac{\beta_\xi}{\lambda x^4}\ln(cF(x))\right]
\eea 
where $x=\phi/\chi$ and $ c$ is a constant.
A nontrivial minimum exists for the field values  $(\phi_0,\chi_0)$ if:
\bea
\label{minimum}
\frac{\partial W}{\partial x} & = &0 \rightarrow cF(x_0) =  \exp(x_0F'(x_0)/4F(x_0))
\nonumber \\
\frac{\partial W}{\partial\phi} & = & 0 \rightarrow 1+ \frac{\beta_\xi}{\lambda x_0^4}\ln(cF(x_0))=0
\eea
Combining gives us one combination of the equations:
\beq
\frac{1}{x_0^3}\frac{F'(x_0)}{F(x_0)} = -\frac{4\lambda}{\beta_\xi}
\eeq
An independent combination of the equations gives us a fine-tuning constraint on $c$.

We can consider the simple case,  $F\equiv F_\xi= 1/x=\chi_0/\phi_0$ and
we thus  find, $x_0 =\phi_0/\chi_0= (\beta_\xi/4\lambda)^{1/4}$.
Note however the consistency condition,
$\ln(cF(x_0))=-\lambda x_0^4 /\beta =-1/4$ requires that 
$c$ is fine tuned as: $c = x_0\exp(-1/4) $.

Once tuned, this not only corresponds to a minimum but also to a zero of the potential,
 i.e. a locus in field evolution of fixed $x_0 =\phi_0/\chi_0$
with no cosmological constant. It is straightforward to consider the more general
case with a fixed point and late time accelerated expansion, 
generalizing the results we found in the previous sections. 

 Including
a  running $\alpha_1$ term and $\alpha_2 \approx$ constant, we have the action:
\begin{eqnarray}
S&=&\int\sqrt{-g}\left\{\frac{1}{2}g^{\mu\nu}\partial_\mu\phi\partial_\nu\phi
+\frac{1}{2}g^{\mu\nu}\partial_\mu\chi\partial_\nu\chi-W(\phi,\chi)\right. \nonumber \\ & &\left.
-\frac{1}{12}\left[1+(\alpha_0-1)F(x)^{\gamma_1}\right] \phi^{2}R 
-\frac{1}{12}{\alpha_2} \phi ^{2}R\right\}\; . 
\end{eqnarray}

\begin{figure}[t]
\includegraphics[width=8.9cm]{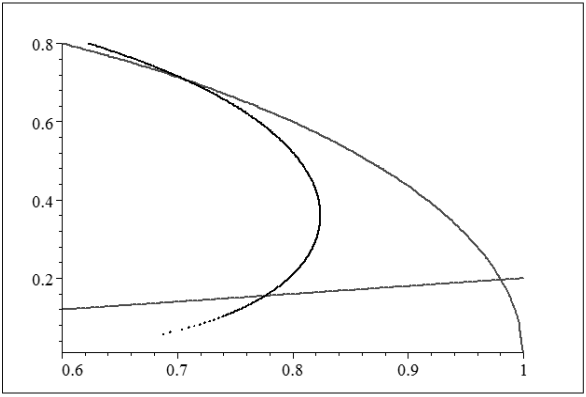} 
\vspace{-2.5cm}
\caption[]{
{We expand
the right-handed quadrant, $(\phi,\chi)>0$ 
where the classical
ellipse has rightmost endpoint at $\chi =1$.
The  ``quantum ellipse" turns back toward
the origin due to the quantum running of $\alpha_1$
where  $\phi$ tracks $\chi$ as $\phi \propto \chi^{\gamma_{\alpha}/\left( 2+\gamma_{1}\right) }$.
The potential flat direction 
is indicated as the nearly horizontal line.}}
\label{xroots3}
\end{figure}

The quantum corrections  deform the ellipse shown in Fig.(2),
arising from the running of the $\alpha_i$ (mainly $\alpha_1$ presently).
 $\gamma_1$ is a parameter
appearing in the $\beta$-function for
$\alpha_1$, and
$\alpha_{1_0}<0$ is an initial value of $\alpha_1$
at the ``scale'' $\phi/\chi=1$. We reiterate that, in the Weyl invariant
framework, one must get used to the notion that
there are no fundamental mass scales anymore, and only
invariant ratios of field VEV's can arise
in scale invariant physical quantities such
as dimensionless couplings like the $\alpha_i$.

Hence, given the fixed value of $K$, we have in
the classical and quantum cases: 
\bea
\label{Kseq1}
\makebox{classical:} \;\;&& 2{K}=\left( 1-\alpha _{1}\right)\phi^2+
(1-\alpha _{2})\chi^2
\nonumber \\
\makebox{quantum:}\;\; && 2{K}=\left( 1-\alpha _{1_0}\right)\phi ^{2} 
\left[F(x)\right] ^{\gamma_{1}}+(1-\alpha _{2})\chi^2
\nonumber \\
&&
\eea
If we now specialise to $F(x)\equiv F_\alpha = x$, we find the differences  illustrate in Fig.(\ref{xroots3}).
In this case, the Planck mass is now given by:
\bea
\label{Kseq2}
\makebox{classical:}\qquad 6M_P^2 &=& -\alpha _{1} \phi ^{2}-\alpha _{2}\chi ^{2}
\nonumber \\
\makebox{quantum:}\qquad 6M_P^2 & =& 
\nonumber \\
& & \!\!\!\!\!\!\!\!\!\!\!\!\!\!\!\!\!\!\!\!	\!\!\!\!\!\!\!\!\!\!\!\!\!\!\!\!\!\!\!\!\!\!\!
-\left(1-\left( 1-\alpha _{1_0}\right)\left( \frac{\phi }{%
\chi }\right)^{\gamma_{1}}\right) \phi ^{2}-\alpha _{2}\chi ^{2}
\eea
At the rightmost end of the ellipse, where $\chi \rightarrow 0$ 
we thus have, approximately:
\bea
\label{Kseq3}
\makebox{classical:}\qquad  & & 2K\approx (1-\alpha _{1})\phi ^{2}
\nonumber \\
\makebox{quantum:}\qquad& &  2K\approx \left( 1-\alpha _{1_0}\right) \phi ^{2}\left( \frac{\phi }{\chi }%
\right) ^{\gamma_{1}}
\eea
The fields ultimately reach the fixed point, the intersection of the
ellipse and the flat direction, and then satisfy the potential minimum constraint $\chi=\epsilon \phi$.

In the classical case the results are simple. We see that $\phi$ is determined by eq.(\ref{Kseq3}),
and likewise, $6M^2_P\approx -\alpha _{1} \phi ^{2}$ follows in  $\chi\rightarrow 0$
limit from eq.(\ref{Kseq2}).
We also have that $\chi=\epsilon \phi$ is determined from the flat direction
of the potential. Hence, $M^2_P=-\alpha _{1}K/3(1-\alpha _{1})$. This
defines the vacuum of the theory, and
the slow-roll migration along the ellipse to the fixed point can generate many e-foldings
of inflation, $N\propto -1/\alpha$ (see Section V).

The quantum case is somewhat different. As $\chi \rightarrow 0$, we see
that the constraint of fixed $K$ and the running of $\alpha_1$ cause
 $\phi$ to track $\chi$:
\beq
\phi \propto \chi^{\gamma_{1}/\left( 2+\gamma_{1}\right) }
\eeq
Hence $\chi$ approaches zero quickly, while $\phi$ also tends
to zero but does so more slowly.   
The Planck mass, however, approaches a constant:
\bea
\makebox{quantum:}\;\; & &  6M_{P}^{2}
\approx  \left( 1-\alpha_{1_0}\right)\phi ^{2} \left( \frac{\phi }{\chi }\right)^{\gamma_1}
\eea
thus we see that at the end of the ellipse with $\chi \rightarrow 0$:
\beq
M_{P}^{2}=\frac{1}{6} \left(\left( 1-\alpha _{10}\right)
\left( \frac{\phi }{\chi }\right) ^{\gamma_{1 }}\right) \phi
^{2} =\frac{1}{3}K
\eeq
While $\phi$ and $\chi$ are becoming smaller, they do not
become zero.  The fixed point, the terminus of inflation,
corresponds to the minimum of the potential in the flat direction, hence
\beq
\chi=\epsilon\phi
\eeq
Combining, this yields the final VEV's
of the fields:
\beq
\phi = M_P\left(\frac{6\epsilon^{\gamma_1}}{1-\alpha_{1_0}}\right)^{1/2}
\qquad \chi =\varsigma \phi
\eeq
It is interesting to speculate about the implications
of this result in realistic models.
The present model supposes only the potential interactions
amongst $\phi$ and $\chi$ and the non-minimal gravitational
interactions. The quantities $\gamma_i$ in the present scheme
are determined by the quartic couplings $\lambda,\delta,\xi$ and
involves mixing induced by $\delta$.  If the only relevant
term was $\lambda$, as in the single scalar model,  we compute  $\gamma_1 = 3\lambda/8\pi^2$.
However, with the flat direction we have $\lambda=-\varsigma^2 \delta$,
and mixing effects in $\gamma_i$ are dominant.  
 In any case, if the potential
coupling contributions to $\gamma_i$ are small
and if they are the only effects, we would have the classical
result,  $\phi=c_0M_P$
with $c_0=\sqrt{6/(1-\alpha_{1_0})}$ of order unity. 

However, other schemes would likely have additional interactions,
including gauge interactions. For example, $\phi$ and $\chi$
could have separate $U(1)_i$ gauge groups and gauge
couplings $(e_1,e_2)$, hence $\gamma_{i}\sim ke_i^2/16\pi^2$.
Moreover, what is relevant is the ``UV'' behaviour of these couplings
\ie, the large $\phi/\chi$ limit, and they
could become large.  
Hence, is possible that in such schemes
$\gamma_1$ can become large, perturbatively 
ranging, perhaps, from $\sim 0.1$ to $1$, and nonperturbatively
even larger.
We thus would have $\phi=c_0M_P\epsilon^{\gamma_{1}/2}$ and: 
 \beq
\phi = c_0M_P^{2/(2+\gamma_1)}\chi^{\gamma_1/(2+\gamma_1)}\qquad \chi=\varsigma \phi
\eeq
If we then identify $\chi$ with the Higgs VEV, $v_H=175$ GeV, then
we determine $\phi=c_0 M$ where $M= 2.6\times 10^{13}$ GeV with $\gamma=1$
and $M= 1.6\times 10^{18}$ GeV with $\gamma=0.1$.
So, it possible that the quantum running
of $\alpha_{1_0}$ plays a role in establishing the grand unification scale, 
identified with the VEV of $\phi$.
Even more extreme, if we identify $\chi$ with the QCD scale, $0.1$ GeV
and allow a nonperturbative at large $\phi/\chi$, $\gamma_1\approx 10$, then we find
$\chi\approx v_H\approx 175$ GeV.  Perhaps $\chi$ could then be
identified with the Higgs boson itself (this would be
 a ``Higgs inflation model'' with a dynamically generated Planck mass),
where $M_P\sim m_H(m_H/\Lambda_{QCD})^\gamma_1$.

{The quantum effects are clearly of great interest.
A detailed study of
the renormalization of this theory and various models is beyond the scope of the
present paper (see \cite{FHR2}). In particular the worked example of the ellipse 
we have presented involves a particular choice of an  ``ansatz''
of $F(x)$, that might be anticipated from full calculation. Full details
 will be presented elsewhere \cite{FHR2}.}

\section{Conclusions}

In the present paper we have discussed  how inflation and
Planck scale generation emerge from a dynamics associated with
global Weyl symmetry and its current, $K_\mu$.  In the pre-inflationary universe,
the scale current density, $K_0$, is driven to zero by general expansion.  
However,   $K_\mu$  has a kernel structure, \ie, $K_\mu =\partial_\mu K$,
and as   $K_0\rightarrow 0$,  the kernel evolves as  $K\rightarrow $ constant.  
This resulting constant $K$ 
defines the scale symmetry breaking, indeed, defines $M_P^2$.
The breaking of scale symmetry is thus determined by  
random initial values of the field VEV's. In addition,
a scale invariant  potential of the theory ultimately determines 
the relative VEV's of the scalar fields contributing to $K$.

This mechanism entails a new form of dynamical scale symmetry breaking
driven by the formation of a nonzero kernel, $K$, as the order
parameter of  scale symmetry breaking. The scale breaking has nothing to do with the
potential in the theory, but is dynamically generated by gravity.
The potential ultimately sculpts the structure of
the vacuum (together with any quantum effects that may distort the
$K$ ellipse).
There is a harmless dilaton associated with the dynamical symmetry breaking.

We  illustrated this phenomenon in  a single scalar field theory, $\phi $, 
with non-minimal coupling to gravity
$\sim -(1/12)\alpha \phi^2 R$, and a $\lambda \phi^4$ potential. 
The theory has a conserved current, $K_\mu=(1-\alpha)\phi \partial_\mu \phi$.
The scale current charge density
dilutes to zero in the pre-inflationary phase
$K_{0}\sim (a(t))^{-3}$. Hence, the kernel, $K=(1-\alpha)\phi^2/2 $, and
the VEV of $\phi$ are driven to a
constant. With $\alpha<0$, this 
induces a positive Planck (mass)$^2$. 
The resulting inflation is eternal.
However, if we allowed for breaking of scale symmetry through quantum loops,
by conventional scale breaking renormalization, 
the resulting trace anomaly would imply that $K_\mu$ is no longer
conserved. Then $\phi$ would relax to zero, and so too the Planck mass.

In multi-scalar-field theories we see that
the generalized $K=\sum_i(1-\alpha_i)\phi_i^2/2$. As this is driven to a constant
by gravity, it
defines an ellipsoidal constraint on the scalar field VEV's, and
the Planck scale is again generated $\propto K$.
An inflationary slow-roll is then associated with the field VEV's migrating along the ellipse,
ultimately flowing to an infra red fixed point. This is shown to be amenable to analytic
treatment, again owing to the Weyl symmetry.
If the potential has a flat direction, which is a ray in field space 
that intersects the ellipse, then the fixed point corresponds
to the potential minimum, and the field VEV's flow to it.
This is associated with a period of rapid reheating and relaxation to the vacuum.
This terminal phase of inflation is similar to standard $\phi^4$ inflation, since the effective
theory is now essentially Einstein gravity with a fixed $M^2_P$.
The vacuum is determined by the intersection
of the flat direction and the ellipse. The final cosmological constant vanishes
by the scale symmetry.

These classical  models illustrates the essential requirement of maintaining the Weyl symmetry,
including quantum effects throughout. Any Weyl breaking effect will
show up as a nonzero divergence in the $K_\mu$ current. Quantum anomalies
will occur with conventional running couplings constants ($\beta$-functions). 
We show that a Weyl invariant
condition can be imposed on renormalized coupling constants to enforce the symmetry
in the renormalized action.  The coupling ``constants'' are then functions of
Weyl invariant quantities. For example, $\lambda$, which previously
ran with $\phi/M$, now runs with the Weyl invariant
function of the fields,
$F_\lambda(\phi, \chi, g_{\mu\nu})$. This preserves
all of the features of the classical  global  Weyl invariant model,
but enforces a constraint on the original $\beta$-functions
that can only be satisfied by introducing field dependent counterterms.
This is similar
to adding the Wess-Zumino-Witten term to a theory as a counterterm
to cancel (or provide) unwanted (or desired) chiral anomalies.  
We will explore detailed calculations
that explicitly exhibit these results elsewhere \cite{FHR2}.

We have  experimented with the anticipated effects
of quantum corrections in a simple ansatz model of the quantum effects.  
Here we see that the ellipse may
be significantly distorted near the intersection with a potential flat direction. 
 The final phase of inflation can
involve a trajectory in which both scalar field VEV's shrink, but subject to 
a constraint that maintains constant $K$, and thus constant $M^2_P$.  If the quantum
effects are large, we may generate multiple hierarchies with possible intriguing
relationships, such as $M_P = M_{GUT} (M_{GUT}/m_{Higgs})^\gamma$.

The Nambu-Goldstone theorem applies in these models, with the dynamical scale
symmetry breaking by nonzero $K$, and there is a dilaton.  We touch upon
some of the properties of the dilaton, with a more detailed discussion of its phenomenology
in a  subsequent work \cite{FHR2}. If the underlying
exact  Weyl scale symmetry (though spontaneously broken via $K$) 
is maintained throughout the theory, then the massless 
dilaton has at most derivative coupling to matter, becomes harmlessly
decoupled, and  any putative  Brans-Dicke constraints 
go away \cite{Fujii}. Again, here it is essential that quantum breaking of 
global Weyl scale symmetry
be suppressed to maintain the decoupling of the dilaton.

An unsolved problem in these schemes is that
 the flat direction generally can exist only for the special case
of a fine-tuned parameter. This has been argued to be enforced in certain cases by a symmetry,
such as in an $SO(1,1)$ invariant potential, $\sim \lambda(\phi^2 -\chi^2)^2$ \cite{Linde}. 
However, there is no such symmetry in
the full theory as, \eg, the $\phi$ and $\chi$ kinetic terms are $O(2)$ invariant, and these 
symmetries will clash in loop order, and the flat direction will be lifted.
If $c$ is not fine-tuned, then we get either
a trivial minimum at $\phi_0=\chi_0 =0$, or a saddle-point.
Hence, a fundamental problem for us is how to naturally maintain flat directions.

Though we haven't discussed it in
detail presently, we expect there are implications
here for novel UV completions of gravity. 
There is an inherent   UV ``softening'' of quantum general relativity
in these schemes since, essentially,
we have no graviton propagator in this theory until the Planck scale forms.
The low energy Einstein gravity is then emergent. The UV completion of
gravity would have to be scale-free and it might
be viewed as a theory that contains only a metric, matter fields with non-minimal
couplings, general covariance, but no stand-alone curvature terms.
The construction of such a theory is beyond the scope of
the present paper. 

Global Weyl invariance may be a veritable and profound constraint on nature.
It hints at intriguing consequences, dramatically including
a dynamical origin of inflation and $M_P$ as a unified phenomenon,
dynamically generated mass hierarchies, including
new effects that involve 
the running to the nonminimal coupling parameters,  and leading
ultimately to a vacuum with (near) zero cosmological constant.

After completing this paper we received a related work
by Kannike, \etal, \cite{Kannike},
who discuss the effect of explicit sources of scale 
invariance breaking
on the stability of the Planck scale with non-minimally coupled scalars, 
including Coleman-Weinberg potentials.
The authors find it challenging
to construct viable models,  lending support to the result here that
Weyl symmetry must be maintained and its breaking can only be spontaneous.

\vskip 0.5 in
\noindent
 {\bf Acknowledgements}

We thank W. Bardeen and J. D. Bjorken for discussions.
Part of this work was done at Fermilab, operated by Fermi Research Alliance, 
LLC under Contract No. DE-AC02-07CH11359 with the United States Department of Energy.

\end{document}